\documentclass[%
 twocolumn,
 amsmath,amssymb,
 aps,prfluids,
 superscriptaddress,nofootinbib,showkeys,
]{revtex4-2}

\usepackage{graphicx}
\usepackage{bm}
\usepackage{xcolor}
\usepackage{array}
\usepackage{booktabs}
\usepackage{siunitx}
\usepackage[T1]{fontenc}
\usepackage[utf8]{inputenc}
\usepackage{hyperref}

\newcolumntype{C}[1]{>{\centering\arraybackslash}p{#1}}
\newcommand{\bs}[1]{\boldsymbol{#1}}
\newcommand{\Sc}{\mathrm{Sc}}
\newcommand{\Rey}{\mathrm{Re}}
\newcommand{\Pe}{\mathrm{Pe}}
\newcommand{\diff}{\mathrm{d}}

\begin{document}

\title{Amphibian water to land transition reveals physical limits of olfaction}

\author{Martin James\textsuperscript{\ddag,*}}
\affiliation{Machine Learning Genoa Center (MaLGa) \& Department of Civil, Chemical and Environmental Engineering, University of Genoa, 16145 Genoa, Italy}
\author{Loranzie S. Rogers\textsuperscript{\ddag}}
\affiliation{Department of Molecular and Cellular Biology, Harvard University, Cambridge, MA 02138, USA}
\author{Moreira Salsman}
\affiliation{Department of Molecular and Cellular Biology, Harvard University, Cambridge, MA 02138, USA}
\author{Francesco Viola}
\affiliation{Gran Sasso Science Institute (GSSI), L'Aquila 67100, Italy}
\affiliation{INFN--Laboratori Nazionali del Gran Sasso, Assergi, Italy}
\author{Nicholas W. Bellono\textsuperscript{\S}}
\affiliation{Department of Molecular and Cellular Biology, Harvard University, Cambridge, MA 02138, USA}
\author{Agnese Seminara\textsuperscript{\S,*}}
\affiliation{Machine Learning Genoa Center (MaLGa) \& Department of Civil, Chemical and Environmental Engineering, University of Genoa, 16145 Genoa, Italy}

\date{September 27, 2026)}

\begin{abstract}
The evolutionary transition from water to land required animals to sense chemicals under fundamentally different physical constraints. Because chemicals diffuse orders of magnitude more slowly in water than in air, aquatic olfaction faces significant challenges in odor capture. Yet, aquatic organisms exhibit strikingly sensitive and rapid chemosensory behaviors. Amphibians represent a powerful model to address this discrepancy because they transition from aquatic to terrestrial environments within a single lifetime. Here, we combine metamorphosis of adult axolotls with asymptotic theory and numerical simulations to show that odorants are captured by diffusion in tens of seconds in water compared with milliseconds in air; yet aquatic axolotls respond to odorants on a much more rapid timescale. Despite inefficiency, we find that proximity to large patches of highly concentrated odor supports sensation in the presence of flow. While aquatic axolotls did not display any respiratory flow, they display behavioral adaptations that generate flow and may even draw in odor through ciliary pumping. Thus, aquatic olfaction is slower and more wasteful than terrestrial olfaction, but behavioral adaptations support olfaction as an effective proximal sense.
\end{abstract}


\maketitle

{\renewcommand{\thefootnote}{}\footnotetext{\textsuperscript{\ddag,\S}These authors contributed equally to this work.\\ \textsuperscript{*}To whom correspondence should be addressed. E-mail: \href{mailto:martin.james@edu.unige.it}{martin.james@edu.unige.it}; \href{mailto:agnese.seminara@unige.it}{agnese.seminara@unige.it}}}

\section{Introduction}
\label{sec:intro}

Physical stimuli reach the sensory organs by travelling through fluids: either air or water. When vertebrates emerged from water and colonized land, profound physiological and morphological adaptations occurred to attune their sensory systems to the physical properties of air. For olfaction, the tight and cohesive arrangement of water molecules dramatically slows the diffusion of small molecules, making diffusive odor capture approximately 10,000 times slower in water than in air. Despite this constraint, the sense of smell enables spectacular homing and navigation behaviors in aquatic animals including predation by sharks~\cite{nosal2016olfaction}, spawning in salmon ~\cite{wisby1954effect} and predator avoidance in crustaceans~\cite{grasso2002lobsters,shabani2008spiny}. How do aquatic organisms speed up odor capture to facilitate behavior? 

\begin{figure}[t!]
\centering
\includegraphics[width=1\columnwidth]
{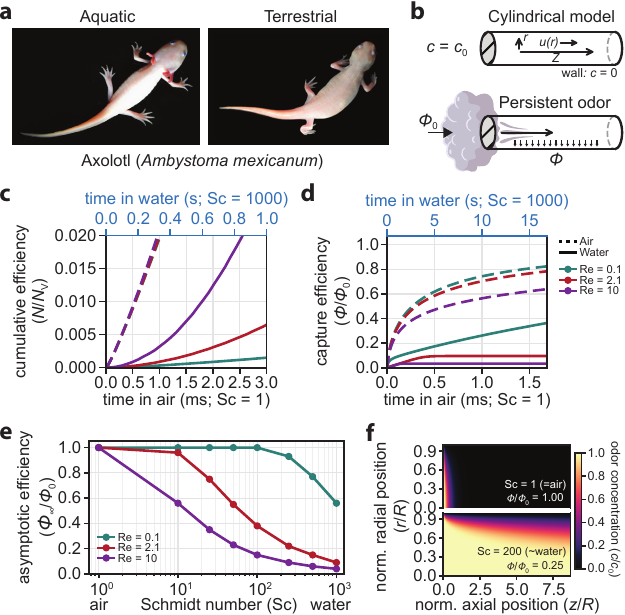}
\caption{Odor is captured much more efficiently by axolotl after transition to land. 
{\bf a}, Laboratory induced transition from aquatic (left) to terrestrial (right) adult \emph{A.~mexicanum} axolotls. 
{\bf b}, Cylindrical model of olfactory chamber, length $L$ and radius $R$ ($R = 0.53$\,mm and $L/R=8.67$ match aquatic morphology). A large, persistent odor patch with concentration $c_0$ enters the nose at velocity $U$, generating a Poiseuille profile $u(r)$ in the chamber. A fraction of the incoming odor reaches the epithelium where it is captured. 
{\bf c}, Cumulative odor captured by the epithelium, $N$, normalized with $N_V=c_0\times \pi R^2L$ in air (dashed) and water (solid) at different Reynolds numbers.
{\bf d}, Capture efficiency, $\phi/\phi_0$, where $\phi=dN/dt$ is the capture rate and $\phi_0=c_0U\pi R^2$ is the rate of incoming odor. Efficiency reaches much larger values and much more quickly in air (bottom time axis) relative to water (top time axis). 
{\bf e}, Asymptotic capture efficiency $\phi_\infty/\phi_0$ as a function of $\Sc$.  
{\bf f}, Odor concentration within the cylinder at $\Sc=1$ (air) and $200$ (approximating water) for $\Rey=2.1$. All results are obtained by solving Eq.~\eqref{eq:cylinder_nondim}, and can be related by rescaling Sc, as they only depend on Pe.
}
\label{fig:cylinder}
\end{figure}

Here, we leverage amphibian metamorphosis as a model to investigate how the physical constraints on chemosensation differ between air and water. Amphibians transition from an aquatic to a terrestrial state in one lifetime, offering a living glimpse into evolutionary transitions between aquatic and terrestrial ecosystems. Under induced metamorphosis, we find that the \emph{Ambystoma mexicanum} olfactory chamber transforms its morphology, making it possible to compare water-like and air-like transport within related anatomical designs (Fig.~\ref{fig:cylinder}a)~\cite{stuelpnagel2005olfactory,rozanski2020macro}. In most vertebrates, odors are sensed within an enclosed olfactory chamber where odor molecules are captured by olfactory receptors that bind odorants and generate neural signals that elicit behavior. Micro-computed tomography ($\mu$CT) scans of the axolotl olfactory chamber reveal that the olfactory chamber’s interior is rather unobstructed. This simple morphology, with the olfactory receptor filled epithelium lining the walls of the chamber, contrasts with chemosensory tissues containing convoluted structure features that improve odor capture through sharp curvatures ~\cite{rygg2013computational} and turbulent mixing~\cite{cox2008hydrodynamic}. In the absence of flow, odor capture by the olfactory epithelium depends on diffusion, with a characteristic timescale of $R^2/D$, where $R$ is the chamber's half-width and $D$ is the odor diffusivity. 
In air, $D\sim$\SI{e-1}{\square\centi\meter\per\second}~\cite{tang2015compilation}, whereas in water  $D\sim$\SI{e-5}{\square\centi\meter\per\second}~\cite{delgado2007molecular}. For the chamber dimensions in our experiments (R $\sim$~0.5 mm) our model predicts that capturing 10\% of an odor pulse by diffusion alone takes over 30 seconds in water but only $\sim$~3 ms in air. Yet, aquatic axolotls behaviorally respond to odorant molecules dissolved in the tank. How do aquatic axolotls achieve sensation?

Our model shows that, despite inefficiency, proximity to large patches of highly concentrated odor supports sensation in the presence of flow. The exact benefits of flow depend on duration of odor exposure and mechanism of flow generation, e.g., active ciliary pumping, respiration or flow in the environment. To test this prediction, we isolated an olfactory receptor-ligand pair and measured the odor flux required to trigger a response~\emph{in vitro}. We show a quantitative match between theory and behavioral experiments with controlled flow-induced odor delivery. The match demonstrates that odorants only trigger a behavioral response at extremely high concentrations \emph{in vivo} due to the fundamental inefficiencies of odor capture in water and relatively simple tissue morphologies. 

Together, these results link physical limits of odor capture in water to behavioral sensitivity. Flow can facilitate sensation, but it requires high external odor concentrations to trigger a response. Thus, our findings suggest that aquatic axolotls appear adapted to use olfaction as a rather wasteful proximal sense. 

\section{Results}
\label{sec:results}
\begin{figure*}[ht]
\centering
\includegraphics[width=1.0\linewidth]
{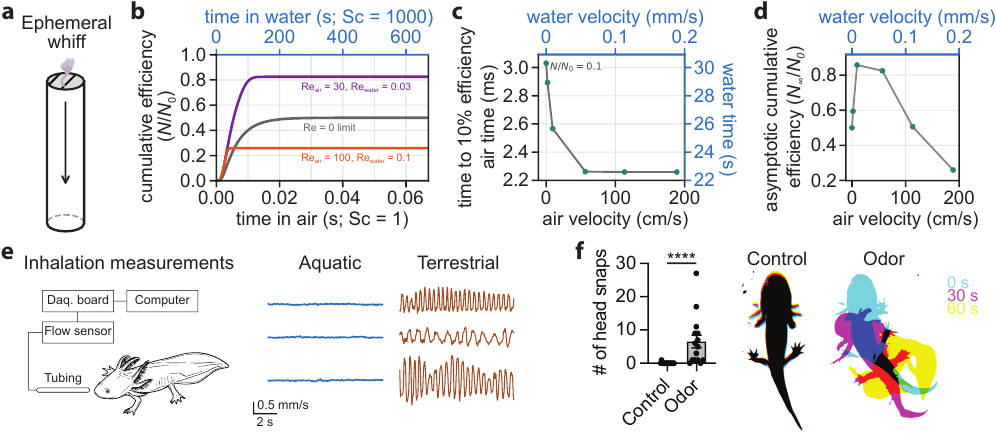}
\caption{Aquatic axolotls respond to prey derived chemicals despite the inefficiency of odor capture in water. 
{\bf a}, An ephemeral whiff containing $N_0$ odor molecules enters the olfactory chamber. 
{\bf b}, Cumulative capture efficiency $N/N_0$ at different $\Rey$ (colors), as a function of time saturates to a plateau, much faster in air (bottom time axis) than in water (top time axis). 
{\bf c}, Time to 10\% efficiency, ($N/N_0 = 0.1$), as a function of flow velocities in air (bottom axis) and water (top axis). Even with fast flow, the time to reach 10\% efficiency remains of order ms in air and tens of seconds in water. 
{\bf d}, Asymptotic capture efficiency $N_\infty/N_0$ as a function of flow in air (lower axis, $\Sc=1$) and water (upper axis, $\Sc=1000$). At the optimal speed  ($\sim $20 cm/s in air and $\sim 0.02$ mm/s in water),  efficiency is above 80\% and plummets for larger speeds. Similar results hold in non-dimensional form (Fig.~\ref{fig:si:pulse} in the Supporting Information).
{\bf e}, Inhalation based flow is observed in terrestrial but not aquatic animals. Left, schematic of inhalation measurements. Right, Flow produced during inhalation by aquatic and terrestrial animals as a function of time. 
{\bf f}, Despite the lack of inhalation, prey-derived odors evoke significant behavioral changes in aquatic animals (prey odor vs. background control, Mann-Whitney U test, U = 18, p < 0.0001). Right, Kymograph showing differences in axolotl movement at 0, 30, and 60 s post-molecule introduction. 
}
\label{fig:pulse}
\end{figure*}
To ask how olfaction is regulated in aquatic versus terrestrial conditions, we conducted laboratory-induced metamorphic transitions in \emph{A.~mexicanum} and studied odor sensing in these respective states (Fig.~\ref{fig:cylinder}a). We first developed a minimal model of the olfactory chamber that incorporates two key processes governing odor capture: advection along the chamber and diffusion toward the odor adsorbing epithelium. We represent the chamber as a cylinder of length $L$ and half-width (or radius) $R$ with air or water flowing into the chamber at a characteristic speed $U$ and generating a Poiseuille velocity profile, $u(r)$ (Fig.~\ref{fig:cylinder}b). Odor molecules are assumed to be immediately captured upon reaching the olfactory receptor lined epithelium, as even sparse receptor coverage yields high odor capture efficiency~\cite{berg2025random}. The normalized odor concentration field is a function of the P\'eclet number $\mathrm{Pe}=UR/D$, which depends on speed $U$, radius $R$ and diffusivity $D$ (Appendix~\ref{app:methods}, Eq.~\eqref{eq:cylinder_nondim}). 
To separate the effects of flow speed and fluid medium, we use $\mathrm{Pe}=\Rey\Sc$, where the Reynolds number, $\Rey=UR/\nu$, depends on flow and the Schmidt number, $\Sc=\nu/D$, represents the fluid medium; $\nu$ is the kinematic viscosity and $D$ is diffusivity. In water, the Schmidt number ($\Sc$) is $\sim$~1000 and $\sim$~1 in air.\\

We first consider a large, persistent odor patch (Fig.~\ref{fig:cylinder}b) that maintains a constant odor concentration, $c_0$, at the entrance of the chamber. As odor is advected through the chamber, it also diffuses laterally and is captured at the walls. From the moment the odor is present, the cumulative amount captured, $N$, by the epithelium builds up, with substantially faster accumulation in air than in water (Fig.~\ref{fig:cylinder}c). The epithelial odor capture rate, $\phi=dN/dt$ (moles/second), also increases initially and then asymptotically saturates to a constant at steady state. Odor capture is both faster and more efficient in air than in water, as seen by the capture rate and the fraction of incoming odor captured, $\phi/\phi_0$, where $\phi_0\sim \pi R^2 U c_0$ is the rate of odor flowing into the chamber (Fig.~\ref{fig:cylinder}d). Because $\phi_0$ increases linearly with flow speed ($U$), and thus with $\Rey$ for a given medium, fast flow delivers more odor to the chamber. As a result, both $N$ and $\phi$ increase with $\Rey$, however odor capture grows slower than delivery ($\phi_0$), causing the efficiency, $\phi/\phi_0$, to \emph{decrease} with $\Rey$. In other words, fast flow increases the total odor molecules captured, but results in even more molecules wasted. At long times, when odor is at equilibrium, the asymptotic efficiency ($\phi_\infty/\phi_0$) can be calculated analytically (Supplementary material Sec.~\ref{sec:si:molecules}) as a function of $\Rey$ and $\Sc$ (through $\Pe$). At high $\Pe$ the adsorbed flux scales as $\phi_\infty\sim \Rey^{1/3}$ (Eq.~\eqref{eq:si:adsorption_limit}). Importantly, asymptotic capture efficiency decreases monotonically with $\Sc$ robustly for all $\Rey$.\\

\begin{figure*}[ht]
\centering
\includegraphics[width=1.0\linewidth]{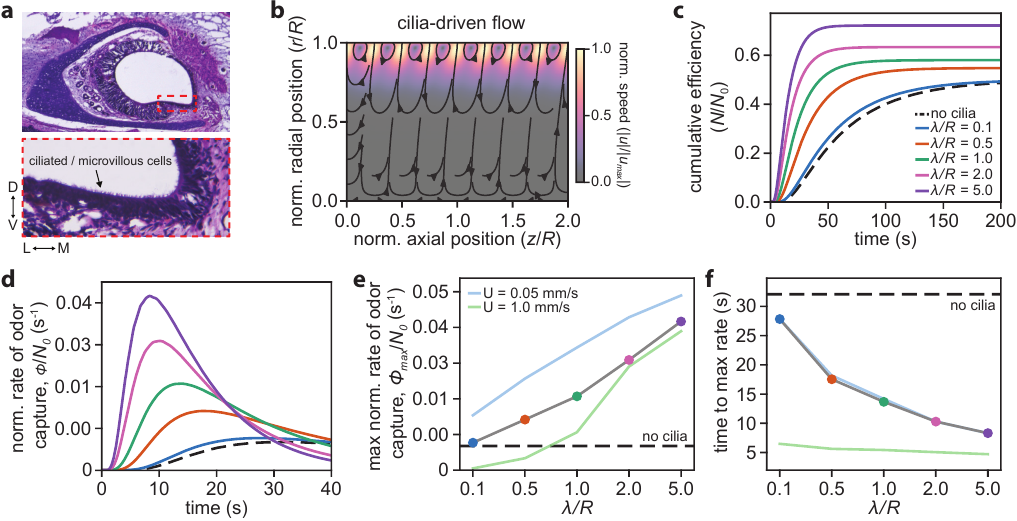}
\caption{Metachronal beating of cilia 
can increase the rate of odor adsorption and maximum odor flux.  
{\bf a}, Aquatic axolotl olfactory chamber is lined with ciliated and microvillous cells. Hematoxylin and eosin staining was used to reveal structural differences in the axolotl head. 
{\bf b}, Streamlines and magnitude of the velocity field due to the metachronal beating of cilia at the top wall. 
{\bf c}, Cumulative capture efficiency $N/N_0$ as a function of time. 
{\bf d}, Normalized rate of odor capture by epithelium as a function of time.
{\bf e}, Maximum rate from {\bf d}.
{\bf f}, Time to reach the peak in {\bf d}. 
In {\bf e} and {\bf f}, the blue and green curves show the corresponding quantities for added mean flows of $U = 0.05$ mm/s and $1$ mm/s, respectively. 
In {\bf c}-{\bf f}: colors represent different values of $\lambda$ and dashed lines the baseline without cilia. All results are shown for capture of ephemeral whiffs in water ($\Sc = 1000$, $\Rey$ = 0).}
\label{fig:cilia}
\end{figure*}

Because odor capture is faster and more efficient in air, odor distribution in the chamber is markedly different for air and water. In air ($\Sc=1$), odor localizes near the entrance of the chamber and is efficiently depleted (Fig.~\ref{fig:cylinder}f, top). In water (larger $\Sc$), radial diffusion is slow and cannot carry the odor from the center of the chamber to the epithelium within the residence time. Therefore, a large proportion of odor molecules remain within the core of the chamber and leave without ever reaching the epithelium and being captured (Fig.~\ref{fig:cylinder}f, bottom).

So far, we have assumed a large persistent odor patch at the entrance of the olfactory chamber. However, odor released from distal olfactory targets travels in randomly fluctuating currents and often reaches the nose in ephemeral whiffs~\cite{celani2014odor}. To mimic this condition, we consider an instantaneous minute odor whiff containing $N_0$ moles of odor located at the center of the inlet (Fig.~\ref{fig:pulse}a). For $\Rey=0$, the process is purely diffusive and half of the odor diffuses into the cylinder while the rest diffuses outside. Because the flow vanishes, for long cylinders with length $L\gg R$, upon waiting long enough, all of those molecules reach the epithelium giving us an asymptotic baseline $N_\infty/N_0=50$\% for the cumulative capture efficiency (Fig.~\ref{fig:pulse}b,d). As the flow rate increases, odor is transported into the cylinder, thus reducing the diffusive loss outside the cylinder and raising asymptotic efficiency above 50\% (Fig.~\ref{fig:pulse}b,d). Furthermore, the increased diffusion to the wall reduces the time to reach a minimum threshold of capture, taking only milliseconds in air but tens of seconds in water (Fig.~\ref{fig:pulse}c).

Interestingly, for an ephemeral whiff, a faster inflow $U$, which may be generated for example by stronger environmental currents, does not always improve efficiency and capture time. There exists an optimal speed,  $U_{opt}$, or $\Rey_{opt}$, in the nondimensional form, that maximizes asymptotic efficiency and speeds up odor capture (Figs.~\ref{fig:pulse}c,d and \ref{fig:si:pulse}). Faster flows decrease asymptotic efficiency and do not further speed up capture. The optimal speed in air is of order $\approx 10 $~cm/s, whereas in water it is only $\approx 0.01\,$mm/s. The difference between air and water occurs because to be at the optimal  $\mathrm{Pe}_{opt}$, $U_{opt}^{air}/U_{opt}^{H20}=D^{air}/D^{H20}$. The value of this optimal $\Rey$ can be easily understood from a scaling argument. The diffusive time to capture is $\sim R^2/D$; the residence time of the molecules within the chamber is $\sim L/U$. If diffusion is slower than advection, odor molecules leave the chamber before reaching the walls, hence we expect maximum efficiency when these two timescales are comparable: $\frac{L}{U}\approx \frac{R^2}{D}$. Hence:
\begin{equation}
\Rey_{\mathrm{opt}}\approx \frac{1}{\Sc}\frac{L}{R}.
\end{equation}
In Appendix~\ref{app:instantaneous}, we refine this argument using the mean absorption location and obtain the corrected estimate
$ \Rey_{\mathrm{opt}}\approx \frac{8}{3}\frac{L}{R}\frac{1}{Sc}$. 
\noindent Note that over the long timescales relevant to water, the rate of odor capture ($\phi$) may be more relevant than the cumulative odor capture ($N$), as sensory neurons are expected to respond to sufficiently large binding rates, rather than to an indefinitely accumulated dose of odor molecules. The behavior of the odor capture rate $\phi=\diff N/\diff t$ mirrors that of the cumulative adsorption. The peak rate is maximized at an intermediate $\Rey$, while the time to reach it decreases with $\Rey$ before saturating (Fig.~\ref{fig:si:pulse_flux}b,c).

Thus, under both persistent odor patches and ephemeral whiffs our cylindrical model confirms the dramatic inefficiency of odor capture in water, which takes $\sim$ tens of seconds, compared to a few milliseconds in air. For a persistent odor patch, faster flow increases odor capture rates, however, many odor molecules escape without reaching the epithelium. For ephemeral whiffs, odor capture by the epithelium is optimal when flow speeds are approximately $\approx 0.01 mm/s$ in water and $\approx 10 cm/s$ in air. To determine if axolotls can produce these optimal flow speeds, we measured air and water flow speeds near the entrance of the nostril using pressure probes. Terrestrial axolotls exhibited consistent breathing patterns with flow speeds on the order $\sim$ mm/s, while aquatic axolotls lacked any consistent respiratory patterns (Fig.~\ref{fig:pulse}e). Yet, aquatic axolotls displayed robust behavioral responses to odors introduced into the tank, snapping their head toward the delivery pipette (Fig.~\ref{fig:pulse}f). 
\begin{figure*}[th]
\centering
\includegraphics[width=0.9\textwidth]{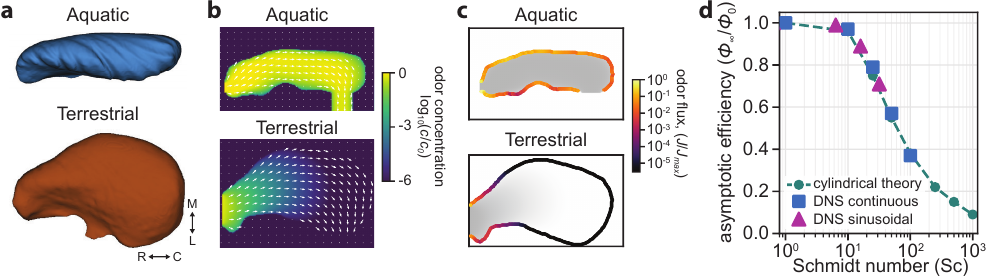}
\caption{
Morphology of the olfactory chamber does not affect capture efficiency, as shown by direct numerical simulation.
{\bf a}, Reconstructed 3d morphologies of olfactory chambers in aquatic and terrestrial axolotls.
{\bf b}, Odor concentration within the olfactory chamber obtained from direct numerical simulations (color coded), and velocity field (white arrows), with a permanent odor patch and $\Rey=2.1$. Within the aquatic morphology at $\Sc=100$ (top), odor concentration remains large over most of the chamber and most of the odor exits the chamber without reaching the epithelium.  
Within the terrestrial morphology 
at $\Sc=1$ (bottom), the odor quickly reaches the epithelium where it is captured, causing a strong depletion within the bulk of the chamber. 
{\bf c}, Flux of odor molecules captured by each portion of the epithelium per unit time, $J=|D\nabla c\cdot\hat{\mathbf{n}}|$, normalized by maximum flux (color) projected onto the visualization plane. Grey shade reproduces the odor concentration (colored in {\bf b}). 
At this $\Rey$, odor is captured across the entire epithelium in water, whereas it is mostly captured near the entrance in air (bottom), consistent with {\bf b}.
{\bf d}, Asymptotic capture efficiency $\phi_\infty/\phi_0$
as a function of $\Sc$ for the aquatic chamber: realistic morphologies do not deviate appreciably from the cylindrical model. 
The cylindrical theory from Fig.~\ref{fig:cylinder} is shown for reference, together with direct numerical simulations for continuous and sinusoidal inhalation. 
}
\label{fig:cfd}
\end{figure*}

Given the sustained behavioral responses despite the lack of observable respiration through the nostrils in aquatic axolotls, we wondered if an active mechanism could overcome this slow and inefficient sense of smell in water. Surfaces lined with motile cilia can generate coordinated flows~\cite{cox2013ciliary,reiten2017motile,raimondi2026cilia}. In particular, the olfactory chambers of aquatic animals, including fishes and amphibians, are known to be lined with motile cilia~\cite{reiten2017motile,ringers2019role,ringers2023novel,cox2013ciliary}, and in zebrafish larvae these cilia have been shown to draw odorants toward the sensory epithelium~\cite{reiten2017motile,ringers2023novel}. Thus, leveraging histological approaches, we next sought to determine if cilia are similarly present in the olfactory chamber of aquatic axolotls. Upon inspection, we found they are indeed lined with cilia (Fig.~\ref{fig:cilia}a). To test whether such cilia can solve the slow adsorption in water, we line the inner wall of the model cylinder with cilia whose tips trace a circle of radius $\varepsilon R$, with $\varepsilon = 0.01$, beating metachronally with wavelength $\lambda$ and frequency $f$. We fix $f=25$\,Hz, consistent with observations in the olfactory pit of zebrafish larvae~\cite{reiten2017motile} and vary the wavelength $\lambda$ in the range $0.1 \leq \frac{\lambda}{R} \leq 5$ (Appendix~\ref{app:cilia}).

The ciliary motion produces a maximum radial velocity $2\pi\varepsilon R  f$ at the wall, and the resulting flow has a penetration depth $\delta$ set by the beat wavelength, $\delta = \lambda/2\pi$ (Eq.~\eqref{eq:cilia_final}). As expected, this motion drives flow toward the epithelium (Fig.~\ref{fig:cilia}b). To quantify its effect, we evaluate capture efficiency in a chamber with cilia-generated flow. Ciliary beating substantially improves odor capture: the cumulative capture efficiency and the odor capture rate both grow with $\lambda$ (Fig.~\ref{fig:cilia}c-d-e). The cilia also speed up capture, reducing the capture time by a factor of 2 to 3 when synchronized over $\lambda\approx R$ or $2R$
(Fig.~\ref{fig:cilia}f). 
Note that ciliary movement may in fact contribute increased speedup if we lift the assumption of axisymmetric flow, which by construction gives zero radial velocity on the axis. In our model an odor pulse on the axis must still cross a diffusive layer to reach the ciliary flow. Non-axisymmetric beating would break this constraint, generating richer flow patterns that could further enhance odor capture. 
Furthermore, the boundary layer decreases if we account for the fact that cilia beating provides a longitudinal flow $U$. A non-zero longitudinal flow $U$ further enhances odor capture by either increasing the maximum capture rate (low speed) or by speeding up time (large speed) (Fig.~\ref{fig:cilia}e-f). 

Together, these results show that flow present in the environment or generated by active ciliary pumping speeds up odor capture. Active ciliary pumping additionally contributes to enhanced lateral transport toward the epithelium, thus improving the efficiency of odor capture. While these mechanisms can lift odor capture above the diffusive limit set by the medium, we note that odor capture in water remains much slower relative to air. 

Next, we wondered whether morphology of the olfactory chamber can speed up odor capture in water.  Therefore, we acquired micro-computed tomography ($\mu$CT) scans of aquatic and terrestrial axolotl olfactory chambers. These scans revealed marked morphological differences primarily in medial-lateral expansions of the olfactory chamber. However, both morphologies revealed chambers were relatively unobstructed throughout (Fig.~\ref{fig:cfd}a). Next, to test the effect of morphology on odor adsorption, we solved Eqs.~\eqref{eq:ns}-\eqref{eq:advection_diffusion} within two distinct domains, corresponding to the morphologies of aquatic and terrestrial chambers. Within the aquatic morphology at the largest Schmidt number accessible through direct numerical simulations ($\Sc=100$; typical values of $\Sc$ in water are $\sim 1000$), odor concentration remains high throughout most of the chamber, thus indicating that a substantial fraction of the odor exits the aquatic chamber without reaching or being absorbed by the epithelium (Fig.~\ref{fig:cfd}b, top). In contrast, the terrestrial chamber in air quickly depletes odor concentrations upon entering  (Fig.~\ref{fig:cfd}b, bottom). 

The full simulations also reveal where odor molecules would hit the olfactory epithelium lining the chamber wall. To monitor where in the epithelium odor is captured, we define the local flux, $J$, of odor captured by the epithelium per unit surface and unit time. The in-plane projection of the flux $J=-D\partial_n c$, shows a non-uniform distribution (Fig.~\ref{fig:cfd}c). Aquatic olfaction captures odor molecules across the entire epithelium; in contrast, terrestrial olfaction concentrates a large flux predominantly near the entrance of the chamber. The uneven regions where odorant molecules hit the epithelium suggest that sensation may increase if olfactory sensory neurons localize strategically closer to the opening for terrestrial animals and throughout the chamber in aquatic animals.
But do these differences depend on the two distinct morphologies, or rather on the fluid environment, i.e., air \emph{vs} water? 

We find that the exact morphology of the chamber does not affect how much or how fast odor is captured. Indeed, the full simulations with the detailed olfactory chamber morphology closely follow the cylindrical model predictions, across a range of Schmidt numbers  (Fig.~\ref{fig:cfd}d). The asymptotic odor capture efficiency is near maximum efficiency under both continuous and sinusoidal inhalation regimes at $\Sc\lesssim 10$, and drops rapidly with larger $\Sc$. Furthermore, continuous and sinusoidal inhalation show quantitative agreement, suggesting flow variations due to sinusoidal inhalation do not increase adsorption. The agreement between the full simulations and the cylindrical theory confirms that the leading-order behavior is controlled by the competition between axial advection and transverse diffusion and morphologies play no significant role. 

\begin{figure}[h]
\centering
\includegraphics[width=1\columnwidth]{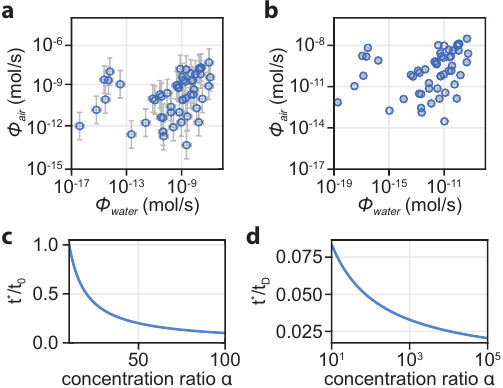}
\caption{Proximity speeds up odor capture. 
{\bf a}, Maximum rate of odor capture $\phi_\infty$ in air ($\phi_{\rm air}$) \emph{vs} water ($\phi_{\rm water}$) for large persistent patch of a saturated odor. Here: $\phi_\infty=\phi_0 \times \frac{\phi_\infty}{\phi_0}$; $\phi_0=c_0\pi R^2U$; $c_0$ are saturated concentrations of a panel of odors and the asymptotic efficiency $\phi_\infty/\phi_0$ is calculated analytically in air eq.~\eqref{eq:si:flux_longtube} and water, eq.~\eqref{eq:si:flux_shorttube}. 
{\bf b}, Maximum rate of odor capture for a saturated, ephemeral odor whiff, computed numerically from the peak in Fig.~\ref{fig:si:pulse_flux}a, $N_0=c_0 V_{\text{whiff}}$ and $V_{\text{whiff}}=4/3\pi R^3$, considering the largest whiff to fit the nose. 
{\bf c,d}, Speed up of odor capture with further concentration. The time $t^*$ to reach 10\% cumulative odor capture when the concentration $c_0$ is multiplied by $\alpha$, decreases relative to the baseline $t_0$. $t^*$ is obtained numerically for odor patches ({\bf c}) and analytically for ephemeral whiffs ({\bf d}).}
\label{fig:molecules}
\end{figure}

All results up to now suggest that olfaction in water is fundamentally inefficient, with many molecules going undetected. However, if axolotls were to use chemosensation near the odor source, odors would reach the nose relatively undiluted and potentially at concentrations near saturation. While odor capture may remain inefficient, i.e., $ \phi\ll\phi_0$, the sheer number of molecules entering the chamber may increase detection ($\phi$) sufficiently to trigger a response. To test how proximity affects the rate of odor capture, we first computed the saturated concentration, $c_0$, of a panel of representative molecules, set by their solubility in water and their volatility in air (Table~\ref{tab:molecules} Supporting Information). For each molecule, $c_0$ sets the baseline both for a persistent odor patch (defined as the rate of incoming odor $\phi_0 \sim \frac{1}{2}c_0 U \pi R^2$) and for an ephemeral whiff (defined by the number of molecules it contains, $N_0=c_0V_{whiff}$). $U$ is taken to vary so that $\Rey \in(0.1,20)$. Using $\phi_0$ and $N_0$, the maximum adsorption rate (moles per unit time) is obtained analytically for the large persistent puff and numerically for the odor whiff. Under these saturated conditions and despite the large inefficiency of olfaction in water, the sheer number of molecules hitting the epithelium spans several orders of magnitude for distinct chemicals  (Fig.~\ref{fig:molecules}a,b). Even for the least soluble/volatile compounds, $\phi_\infty \sim $~100 000 molecules per second reach the epithelium, both in air and in water. 
\\
At the same time, increased concentration also speeds up the time to reach set thresholds. We choose a (arbitrary) threshold for sensation corresponding to $N_t/N_0=10\%$. 
For the persistent patch, from simulations shown in Fig.~\ref{fig:cylinder}, we identify the time $t_0$ to reach the sensation threshold $N_t$ for a reference concentration $\bar{c}$. We then consider a more concentrated source $c_\alpha=\alpha \bar c$, and find that the time to reach the same threshold $N_t$ decreases considerably as we increase $\alpha$  (Fig.~\ref{fig:molecules}c).  For the odor whiff, the speedup can be calculated analytically and it is slower, namely logarithmic (Fig.~\ref{fig:molecules}d).
At times much shorter than the diffusion time, the cumulative odor captured has the closed form (see Sec.~\ref{sec:si:short_time_limit})
\begin{equation}
    N \approx 2N_0\, e^{-t_D/(4t)}, \qquad t_D = \frac{R^2}{D},
    \label{eq:adsorption_short_time}
\end{equation}
where $t_D=R^2/D$ is the diffusion time. Setting $N(t^*) = N_t$ gives the detection time
\begin{equation}
    \frac{t^*}{t_D} = \frac{1}{4\log 2\alpha}, \qquad \alpha \equiv \frac{N_0}{N_t}.
    \label{eq:detection_time}
\end{equation}
Equation~\eqref{eq:detection_time} holds while $t^* \ll t_D$, which is reasonable for aquatic axolotls where $t_D\approx 250$~s.

\begin{figure*}[t]
\centering
\includegraphics[width=0.9\textwidth]{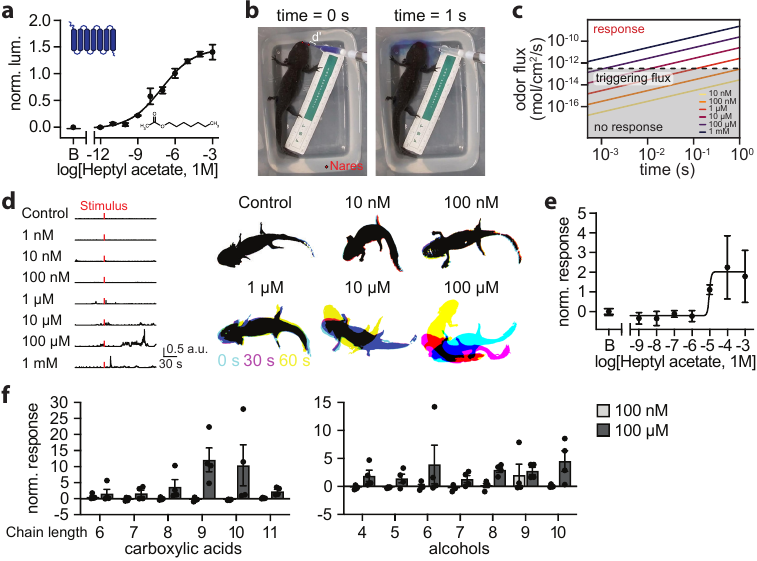}
\caption{
Aquatic axolotls can use olfaction as a proximal sense. {\bf a}, Heterologous expression of an axolotl olfactory receptor in response to heptyl acetate (n = 3 measurements, median effective concentration  $c_{50}$ = 83.6 nM). {\bf b}, Video stills of behavioral measurements in response to 1 mL heptyl acetate at variable concentrations. A blue dye was added to visualize the jet that delivers odor to the nostril. 
{\bf c}, Predicted flux of odorant hitting the epithelium as a function of time. Only concentrations above $10 \mu$M are expected to trigger a response. 
{\bf d}, Only high concentrations of heptyl acetate trigger behavioral responses in aquatic axolotls. Left, activity of axolotls pre- and post-molecule introduction. Right, Kymographs illustrating axolotl movement at 0, 30, and 60 s post-molecule introduction at varying concentrations. 
{\bf e}, Behavioral dose-response curves of axolotls in response to heptyl acetate (n = 3 animals), median effective (behavioral) concentration (EC50) = 9.6~$\mu$M. Response levels during molecule stimulation were baseline subtracted. 
{\bf f}, Across a range of molecules, axolotls are more likely to respond to high odor concentrations. Response levels during molecule stimulation were baseline subtracted 
(n = 4 animals).
}
\label{fig:experimental}
\end{figure*}
\noindent For a 100 fold increase in concentration, $\alpha=100$, the detection time falls by a factor of 10 for the permanent patch and a factor of three for the ephemeral whiff. Similar conclusions are obtained with a threshold on the rate of odor capture, $\phi$, rather than the cumulated odor capture, $N$ (Sec.~\ref{sec:si:timescale_adsorption_limits} in the Supporting Information). Thus, proximity can considerably speed up the process that leads to sensation.

Considering these findings, we next designed a set of experiments to test whether aquatic olfaction functions as a proximal sense. We first needed to understand the sensitivity of the axolotl nose to specific molecules. To do so, we heterologously expressed the axolotl olfactory receptor OR2AT4 in cultured cell lines and measured its sensitivity to heptyl acetate, which revealed a half-maximal effective concentration of approximately 80 nM (Fig.~\ref{fig:experimental}a). Leveraging these results, we then modeled how many odor molecules reached the receptors per unit area per unit time $J_{50}=2c_{50}\sqrt{D/\pi t}$ (Eq.~\eqref{eq:flux_trigger}, Sec.~\ref{sec:J50}), with time $t$ to deliver the ligand set at approximately 1s. We term this value as the triggering flux ($J_{50}$) and assume that odor detection is achieved only when the epithelial odor capture flux, $J$, matches or exceeds this value. To test this hypothesis, we delivered 1~mL heptyl acetate approximately 1~cm from the snout of aquatic adults. Delivery lasted approximately 1s and produced a small jet that generated a persistent odor patch of relatively undiluted odor, with flow speeds at the entrance of the nose on the order of cm/s (Fig.~\ref{fig:experimental}b). Under these conditions, our model predicts that odor capture increases during delivery then plummets as the jet dissipates and the odor is diluted. Because only a small fraction of the molecules are captured by the epithelium, theory subsequently predicts that behavioral responses would require heptyl acetate at concentrations above 10 $\mu M$ (Fig.~\ref{fig:experimental}c, Supplementary Information, Sec.~\ref{sec:model_behavior}). Indeed, behavioral experiments support this prediction, with animals reliably responding to concentrations >10 $\mu$M (behavioral $c_{50} = 9.6 \mu M$; Fig.~\ref{fig:experimental}d-e). Moreover, responses initiated as early as approximately $100\,$ms, confirming that the axolotls can detect odor during delivery. When stimulated with a panel of other odorants, animals consistently responded during odor delivery at 100$\,\mu M$, but not at 100$\,nM$ (Fig.~\ref{fig:experimental}f), demonstrating that aquatic axolotls indeed use olfaction as a proximal sense. 

\section{Discussion}
\label{sec:discussion}

Our experiments and theory suggest that the aquatic axolotl, \emph{A.~mexicanum}, uses olfaction as a proximal sense, responding only to odors near saturation. Conversely, as they undergo metamorphosis and transition to land, we predict that these animals undergo a sudden expansion of their sense of smell driven by physics, mirroring the visual range expansion that occurred as animals first transitioned to land~\cite{maciver2017massive}.

Axolotls are adapted for lentic water habitats~\cite{jiang2017analysis} hence water currents in their natural environment are unlikely to support these behaviors. Moreover, we did not find any reliable respiration pattern. Intriguingly, males use tail movement to actively deliver pheromones to the female for mating~\cite{maex2016courtship}, supporting our results that specific olfactory driven behavior requires an active generation of flow. Beyond behavioral adaptations, active ciliary pumping represents a powerful potential mechanism to make odor capture less wasteful in aquatic organisms. Our model shows that cilia of size $\sim1\%$ of the radius, beating in metachronal waves, can substantially increase capture efficiency and odor flux toward the epithelium if they synchronize over millimeters. However, whether axolotl olfactory cilia actively generate flow is unknown. A separate, spatial feature of our simulations is that odor delivery to the epithelium is heterogeneous and differs between the aquatic and terrestrial chambers. To what extent the distribution of sensory neurons may be adapted to better exploit heterogeneous odor capture remains to be established.

In the absence of active mechanisms, the fluid dynamics of passive odor capture in aquatic axolotls is inefficient. Flow enhances odor capture, but makes the process even more wasteful. For example, for persistent odor patches at high water speeds, more odor flows into the chamber ($\phi_0$) and is captured by the epithelia ($\phi$), which can lead to sensation and trigger a behavioral response as in our experiments. But $\phi$ does not increase quite as much as $\phi_0$, thus although the sensory epithelium captures more odor molecules, an even larger proportion of the incoming odor goes undetected. For ephemeral whiffs, the flow, or $\Rey$ more precisely, sets the ceiling on how much odor can be captured. Flow improves efficiency only up to an optimal value, beyond which efficiency plummets, as molecules exit the chamber before reaching the walls. In unobstructed chambers, no passive adjustment of the flow in the absence of turbulent mixing escapes this trade-off.

The limitations of aquatic olfaction follow from fundamental fluid dynamics and are not specific to axolotls. A possible solution to this fundamental inefficiency is proximal chemosensation, but this remains rather wasteful, as most odor molecules entering the nose leave without ever reaching the olfactory epithelium. When in evolution did chemosensation become a fully functional distal sense? Amphibians evolved from lobe-finned fishes more than 300 million years ago.  Cartilaginous and ray-finned fish, including salmon and sharks clearly developed adaptations to leverage the full potential of olfaction as a distal sense. In addition to ciliary pumping discussed here, various other mechanisms may contribute to speeding up odor capture, including, e.g., recirculating flow causing temporary retention of water to increase the residence time of odor within the chamber~\cite{cox2008hydrodynamic}; active deformation of the chamber wall or jaw protrusion to increase near-wall transport~\cite{cox2008hydrodynamic,nevitt1991fish}; active motion of the organism itself, which increases ventilation and introduces fluctuations and turbulent mixing~\cite{garwood2019olfactory}. What adaptations first facilitated this transition and when did they emerge remains a fascinating direction for future work. 

\section*{Author Contributions}
MJ, LSR, NWB and AS contributed to the design of the research. MJ and AS developed the theory and modeled the experiments. MJ conducted the simulations and analyzed the resulting data. LSR and MS conducted the experiments and analyzed the corresponding data. FV wrote the DNS code, which MJ adapted for the present study. All authors contributed to the interpretation of the results. MJ, AS, LSR and NWB wrote the manuscript with input from coauthors.

\section*{Data Availability}
All data supporting the findings of this manuscript will be made available in public repositories upon publication of the corresponding article. The code used for the full geometry direct numerical simulations is available at \href{https://gitlab.com/gssi-fluids/ib-flows}{https://gitlab.com/gssi-fluids/ib-flows}.

\begin{acknowledgments}
This research was supported by grants to: AS from the European Research Council under the European Union Horizon 2020 research and innovation programme (grant agreement number 101002724 RIDING) and the National Institutes of Health under award number R01DC018789; NWB from  the National Institutes of Health under award number R35GM142697 and Schmidt Sciences, LLC; and a Helen Hay Whitney Fellowship, Burroughs Wellcome Fund, and HHMI Hanna Gray Fellowship to LSR. This work represents only the views of the authors; the European
Research Council Executive Agency and the other funding agencies are not responsible for any use that may be made of the information it contains.
\end{acknowledgments}

\makeatletter
\let\saved@hangfrom@section\@hangfrom@section
\let\saved@sectioncntformat\@sectioncntformat
\makeatother

\appendix

\section{Methods}
\label{app:methods}

\noindent \textbf{Animals.}\\

\noindent Axolotls (\emph{Ambystoma mexicanum}) were raised from eggs at the Ambystoma Genetic Stock Center (University of Kentucky). At Harvard University, animals were individually housed in tanks containing Holtfreter’s solution and kept on a 12 hr light/dark cycle. Animals were fed pelleted food every other day. All experiments were conducted following the guidelines of the Harvard Animal Care and Use Committee.\\

\noindent \textbf{Governing equations}\\

\noindent We consider odor transport in an incompressible velocity field $\bs{u}(\bs{x},t)$. The fluid motion is described by
\begin{align}
\frac{\partial \bs{u}}{\partial t}+\left(\bs{u}\cdot\nabla\right)\bs{u} &=-\frac{1}{\rho}\nabla p+\nu\nabla^2\bs{u},\label{eq:ns}\\
\nabla\cdot\bs{u} &=0, \nonumber
\end{align}
where $p$ is pressure, $\rho$ is density and $\nu$ is kinematic viscosity. The odor concentration $c(\mathbf{x},t)$ obeys
\begin{equation}
\frac{\partial c}{\partial t}+\bs{u}\cdot\nabla c=D\nabla^2 c,
\label{eq:advection_diffusion}
\end{equation}
where $D$ is the molecular diffusivity. The Schmidt number is $\Sc=\nu/D$. We consider two odor sources. For a persistent patch of odor, the inlet concentration is fixed at $c_0$ at all times. For the ephemeral whiff, we use an instantaneous odor pulse at the inlet at time 0, $c(\mathbf{x},t=0)=N_0\delta(\mathbf{x})$. The chamber wall is treated as a perfectly absorbing surface, $c=0$. The capture rate can be computed from the diffusive flux into the wall,
\begin{align}
\phi(t)=\int_{\Gamma_w}D\,\left|\nabla c\cdot\hat{\mathbf{n}}\right|\,\diff S, 
\label{eq:wallflux}
\end{align}
where $\Gamma_w$ is the absorbing wall and $\hat{\mathbf{n}}$ is the wall normal. 
When the inlet concentration is fixed at $c_0$, we normalize $\phi$ with the net incoming rate:
\begin{equation}
\phi_0=\int_{\Gamma_{\rm in}} \left[ c_0\,\mathbf{u}\cdot\hat{\mathbf{n}}_{\rm in}-D\nabla c\cdot {{\mathbf{n}}_{\rm in}}\right]\,\diff S,
\label{eq:incoming_flux}
\end{equation}
where $\mathbf{n}_{\rm in}$ is the normal to the inlet, so that $\phi/\phi_0$ is the capture efficiency, i.e.~the fraction of incoming odorant molecules captured by the walls. 
Integrating $\phi$ over time gives the cumulative odor 
\begin{equation}
N(T)=\int_0^T\phi\,\diff t
\label{eq:odor_total}
\end{equation}
captured by the epithelium up to time $T$. For the permanent odor patch, $N$ is normalized using the constant
\begin{equation}
N_V=c_0\pi R^2L.
\label{eq:odor_total_initial}
\end{equation}
For the instantaneous odor pulse, $N$ is normalized using the total amount of odor instantaneously released by the whiff 
\begin{equation}
N_0=\int_V \left.c\right|_{t=0}\,\diff V,
\end{equation}
where $V$ is the domain volume. $N/N_0$ is the cumulative capture efficiency, i.e.~the fraction of odor molecules contained in the whiff that are captured up to time $t$.

For the persistent odor patch in the steady state, when the advective flux into the chamber is much larger than the diffusive flux, the odor flux can be computed from the difference between the incoming and outgoing fluxes,
\begin{align}
\phi_\infty= \int_{\Gamma_{\rm in}}c_0\,\mathbf{u}\cdot\hat{\mathbf{n}}_{\rm in}\,\diff S - \int_{\Gamma_{\rm out}}c\,\mathbf{u}\cdot\hat{\mathbf{n}}_{\rm out}\,\diff S.
\label{eq:wallflux_steady}
\end{align}
\noindent Asymptotic expressions for $\phi_\infty$ are available in the limit of short and long cylindrical tube (Supplementary material Sec.~\ref{sec:si:molecules}).\\
At finite time, the time dependent rate of capture $\phi(t)$ is obtained numerically, as explained below.\\

\vspace{0.2in}
\noindent \textbf{Cylindrical approximation.}\\

\noindent In the first part, we approximate the chamber as a cylinder of radius $R$ and length $L$ with an absorbing wall. The axial coordinate is $z$, the radial coordinate is $r$ and the axial velocity is assumed to follow a Poiseuille profile,
\begin{equation}
u_z(r)=U\left(1-\frac{r^2}{R^2}\right).
\label{eq:poiseuille}
\end{equation}
Thus the concentration satisfies~\cite{barrera2016graetz,leal2007advanced}
\begin{equation}
\frac{\partial c}{\partial t}+U\left(1-\frac{r^2}{R^2}\right)\frac{\partial c}{\partial z}=D\left[\frac{1}{r}\frac{\partial}{\partial r}\left(r\frac{\partial c}{\partial r}\right)+\frac{\partial^2 c}{\partial z^2}\right].
\label{eq:cylinder_dimensional}
\end{equation}
With $\theta=c/c_0$, $Y=r/R$, $T= tD/R^2$, $Z=z/R$ and Reynolds number $\Rey=UR/\nu$, Eq.~\eqref{eq:cylinder_dimensional} becomes,
\begin{equation}
    \frac{\partial \theta}{\partial T}
+
\Rey \Sc(1-Y^2)\frac{\partial \theta} {\partial Z}
=
\frac{1}{Y}\frac{\partial}{\partial Y}\left(Y\frac{\partial \theta}{\partial Y}\right)
+
\frac{\partial^2 \theta}{\partial Z^2}.
\label{eq:cylinder_nondim}
\end{equation}
The relevant nondimensional parameter in this system is thus Pe = $\Rey\Sc$. In the main text, we often treat $\Rey$ and $\Sc$ separately to distinguish the effects of changing fluid medium from changing the flow. To obtain results presented in Fig.~\ref{fig:cylinder}, we solve Eq.~\eqref{eq:cylinder_nondim} with $\theta(Y,0)=1$, $\theta(1,Z)=0$ and $\partial_Y\theta(0,Z)=0$. Note that in the advective regime, when axial diffusion can be neglected, this is the classic Graetz problem~\cite{brown1960heat,shah2014laminar}. The results for an instantaneous odor pulse are evaluated by placing a Gaussian odor pulse of $\sigma = 0.1R$ at the center of a cylinder of length $2L$. Adsorption is evaluated in the downstream half of the cylinder. For numerically solving the cylindrical problem (Eq.~\eqref{eq:cylinder_nondim}), we use a finite volume method. Unless otherwise specified, we choose $R = 0.53$\,mm and $L/R=8.67$ to be consistent with the aquatic axolotl geometry. The radius is obtained by averaging the equivalent radius of the chamber cross-section at five axial positions, and the length is the centerline distance from the inlet to the outlet.

The capture rate in the cylinder is evaluated, following Eq.~\eqref{eq:wallflux}, as
\begin{equation}
\phi = -\int_0^L 2\pi R D\left. \frac{\partial c}{\partial r}\right|_{r=R} \diff z.
\label{eq:efficiency_cyl}
\end{equation}
For the persistent odor patch with $c_0$ at the inlet, the capture efficiency $\phi/\phi_0$ becomes (Eq.~\eqref{eq:incoming_flux})
\begin{align}
\frac{\phi}{\phi_0} &=&& -\frac{\int_0^L 2\pi R D\left. \partial_r c\right|_{r=R} \diff z}{c_0\int_0^R2\pi r u_z(r)\diff r - \int_0^R2\pi Dr\partial_zc|_{z=0}\diff r}.
\label{eq:adsorption_efficiency}
\end{align}
\noindent and this equation is evaluated at all times from the numerical solution for $\theta$~\eqref{eq:cylinder_nondim} (Fig.~\ref{fig:cylinder}c-e).\\

\vspace{0.2in}
\noindent \textbf{Full-geometry direct numerical simulations.}\\

\noindent For the full numerical simulations, we use 3D reconstructions of the aquatic and terrestrial geometries. The Navier-Stokes equations are solved by three-dimensional direct numerical simulation using a semi-implicit second-order finite-difference scheme~\cite{viola2020fluid,vagnoli2026ib}. The chamber walls are imposed as no-slip, absorbing boundaries through an immersed boundary method~\cite{viola_2025}. Continuous inhalation is imposed by a steady inlet jet. For sinusoidal inhalation, the inlet speed is varied periodically with a period of $10\,\mathrm{s}$. The outlet is treated as an open boundary embedded in a larger computational domain. The odor concentration is fixed at the inlet, with an outflow condition at the outlet. The mean velocities normal to the midplane cross section are  1.7 mm/s and 7.2 mm/s for the aquatic and terrestrial chambers, respectively. We choose a nonzero inhalation speed in water, despite experiments showing negligible inhalation (Fig.~\ref{fig:pulse}e), to get an upper bound on odor capture rate in water. For continuous inhalation, the rate of wall capture $\phi_\infty$ is evaluated using Eq.~\eqref{eq:wallflux_steady} after the solution reached a statistically stationary state (Fig.~\ref{fig:cfd}d). For sinusoidal inhalation, wall capture is evaluated using the average of Eq.~\eqref{eq:wallflux_steady} over one period after discarding the first ten cycles. 

\vspace{0.2in}
\noindent \textbf{CT scans of axolotl olfactory chambers.}\\

\noindent 3D reconstructions of axolotl olfactory cavities were generated using animals that were deeply anesthetized with 1\% w/v MS-222 (Syncaine; Syndel) and euthanized by decapitation. Heads were fixed overnight in 4\% paraformaldehyde in phosphate buffered saline (PBS) and then transferred to 1\% phosphotungstic acid in 70\% ethanol and incubated for 14 days at 4$^\circ$C in complete darkness. Following incubation, samples were rinsed with PBS and secured with cheesecloth in a 50 mL conical tube with a small amount of PBS at the bottom to prevent samples from drying out. Samples were individually scanned on a SkyScan 1273 micro-computed tomography system (Bruker; voltage = 52 kV; current = 120 $\mu$A; pixel size = 9.00 $\mu$m), with a rotation step of 0.11$^\circ$. Reconstruction was performed using NRecon (v. 2.1.0.1) before individual scan files were combined into single image stacks using ImageJ. Olfactory cavities were reconstructed in 3D Slicer~\cite{fedorov20123d} using threshold-based selection, manually refined with the scissor and eraser tools to remove extraneous features.

\vspace{0.2in}
\noindent \textbf{Inhalation speed.}\\

\noindent Measurements (Fig.~\ref{fig:pulse}e) were collected using an airflow sensor (Honeywell, AWM3100V) attached to Silastic tubing (length: 5\,cm). The signal was digitized using a Digidata 1550B digitizer (Molecular Devices) using ClampEx software (Molecular Devices). Using a custom-written MATLAB (v. 2025b, The MathWorks, Inc.) script, sensor voltage values were then converted to airflow using the manufacturer's calibration table. To represent bidirectional inhalation and exhalation, we mirrored the calibration curve, yielding flow as a function of time in standard cubic centimeters per minute (sccm). Using the Silastic tube's inner diameter (3\,mm), we then converted flow to average velocity in cm/s by dividing the flow by the tubing’s cross-sectional area. To directly compare aquatic and terrestrial animal inhalations, we calculated peak-to-peak flow and estimated breathing rate from the inhalation-dominant FFT peak, while treating very low-amplitude aquatic traces as no inhalation. 

\vspace{0.2in}
\noindent \textbf{Behavior.}\\

\noindent Aquatic axolotls were placed in tanks containing 1 L of Holtfreter’s solution and allowed to acclimate for 10 min before trials in response to either prey extract or single molecules. Crude prey extract was prepared by homogenizing 10 g of blackworms (Eastern Aquatics, Inc.) and delivered to animals in 3 mL boluses (Fig.~\ref{fig:pulse}f). Single molecules were diluted to the desired test concentration in Holtfreter’s solution containing food coloring to visualize delivery and tested in 1 mL aliquots. Control (Holtfreter’s solution) volumes matched those of the corresponding stimulus. For all trials, behavior was recorded using a GoPro camera (GoPro, Inc.) positioned overhead. Head snaps following prey extract delivery were quantified using BORIS~\cite{friard2016boris}. For single molecule trials (Fig.~\ref{fig:experimental}), a Gaussian blur filter was applied to videos before thresholding and binarization to extract whole body features. Activity levels throughout trials were then acquired using the kymograph plugin (FIJI)~\cite{fiji}.

\vspace{0.2in}
\noindent \textbf{Cells and dose response curves for heptyl acetate.}\\

\noindent Modified HEK293T cells (Hana3A; from the Matsunami Laboratory, Duke University, USA) were cultured in MEM (Corning) supplemented with 10\% FBS (Gibco), 0.5\% penicillin–streptomycin (Gibco) and 0.5\% amphotericin B (Gibco) at 37$\circ$C and 5\% CO$_2$. Real-time cAMP responses downstream of OR activation were measured using the GloSensor cAMP assay (Promega), as previously described~\cite{billesbolle2023structural}. Two days prior to testing, one-fifth of the cells from a confluent 10 cm plate were seeded into poly-d-lysine coated 96-well plates (Corning). The next day, each 96-well plate was transfected with 10 $\mu$g pGloSensor-20F plasmid (Promega), 5 $\mu$g of RTP1S plasmid and 75 $\mu$g of rho-tagged axolotl OR2AT4 in the pCI mammalian expression vector (Promega) using 20 $\mu$L Lipofectamine 2000 (Invitrogen) in MEM containing 10\% FBS. Approximately 24 h later, plate wells were loaded with 25 $\mu$l of GloSensor substrate (Promega) and incubated for 2.5 h in the dark at room temperature. Agonists, which were reconstituted in HBSS medium (Gibco) at a concentration of 1M, were diluted to 2x the desired test concentration in HBSS media (Gibco) supplemented with copper (30 $\mu$M CuCl$_2$; Sigma-Aldrich) and 2 mM l-glutamine (Gibco) and pH adjusted to 7.0 with a 150 mM solution of sodium hydroxide (Sigma-Aldrich). After agonist addition to each well (25~$\mu$l), luminescence was monitored for 10 cycles using a Synergy, Neo2 plate reader (BioTek). Luminescence was normalized to an empty vector negative control, and axolotl OR2AT4 responses were quantified by calculating the area under the curve and averaging the response across all 10 cycles. Gene cloning and synthesis were performed by GenScript. When needed, plasmids were transformed in Stellar competent cells (TaKaRa), grown in liquid culture, and purified using a Maxiprep kit (ZymoPURE).

\vspace{0.2in}
\noindent \textbf{Quantification and statistical analyses.}\\

\noindent Data were analyzed with custom Python or MATLAB (v. 2025b, The MathWorks, Inc.) scripts or using Prism GraphPad and represented as mean ± SEM. Data were considered significant if p < 0.05 using unpaired two-tailed Student’s t tests or one-way ANOVAs. 

\section{Optimal flow for maximum wall capture of an ephemeral whiff inside a cylinder.}
\label{app:instantaneous}

\noindent We are interested in evaluating the optimal flow rate to maximize the adsorption of an odor whiff located at the center of the inlet. Eq.~\eqref{eq:cylinder_nondim} is our starting point. Our initial condition is a point source at the center of the inlet, given by
\begin{equation}
    \theta(Y,Z,T=0) = \frac{\delta(Y)\delta(Z)}{2\pi Y}
\end{equation}
such that the total odor integrates to unity. Let us find $\mathrm{Pe}=\Rey\Sc$ that maximizes odor capture onto the cylinder wall. For this, we can optimize the wall hit position $Z_w$ of an odor particle that starts at the inlet at time $T=0$ so that the mean wall hit position lies at the midpoint of the cylinder (i.e., $\mathbb E\left[Z_W\right] = \frac{L}{2R}$).
To calculate this, let $T_w$ be the first time when an odor molecule hits the wall.
\begin{equation}
    Z_w = \Rey\Sc\int_0^{T_w}(1-Y_t^2)dt + B_{T_w},
\end{equation}
where $B_{T_w}$ is the axial stochastic motion, whose expectation is zero.

To evaluate the expectation of the above function, we note that $\mathbb E\left[T_w\right]=\frac{1}{4}$ and $\mathbb E\left[\int_0^{T_w}(1-Y_t^2)dt\right]=\frac{3}{16}$~\cite{koralov2007theory}, resulting in the mean adsorption location of
\begin{equation}
    \mathbb E\left[Z_w\right] = \frac{3}{16}\Rey\Sc.
\end{equation}
Since we want this to be near the middle of the cylinder, we get
\begin{equation}
    \Rey\Sc=\frac{8}{3}\frac{L}{R}.
\end{equation}

\section{Cilia-driven flow and wall capture}
\label{app:cilia}

Inspired by previous works on hydrodynamics of ciliated surfaces~\cite{brennen1974oscillating,blake1971spherical}, we derive the equations for the flow and adsorption inside a ciliated chamber as follows. Let us again approximate the olfactory chamber as a cylinder of radius $R$ and length $L$. Assume that the inner surface of the cylinder is lined with cilia whose tips trace circular orbits of radius $\varepsilon R$, beating metachronally with wavenumber $k$ and angular frequency $\omega_f$. When $\varepsilon\ll 1$, we can approximate the wall boundary condition as the cilia tip velocity $u_r|_{r=R}=\varepsilon R\,\omega_f\cos(kz-\omega_f t)$ and $u_z|_{r=R}=\varepsilon R\,\omega_f\sin(kz-\omega_f t)$. 

The unsteady Stokes equation for the fluid flow inside the cylinder can be written as~\cite{happel2012low}
\begin{equation}
    \partial_t\!\left(E^2\psi\right)=\nu\,E^4\psi,
    \label{eq:stream_function}
\end{equation}
where $\psi$ is the stream function such that $u_z = \frac{1}{r}\partial_r\psi$ and $u_r = -\frac{1}{r}\partial_z\psi$, and $E^2\equiv\partial_r^2-\frac1r\partial_r+\partial_z^2$. We seek solutions with the same space and time dependence as the wall forcing
\begin{equation}
\psi(r,z,t)=\Re\left[\tilde\psi(r)\,\mathrm{e}^{\,i(kz-\omega_f t)}\right].
\label{eq:psi_solution}
\end{equation}
Substituting this into Eq.~\eqref{eq:stream_function} and defining
$\mathsf{L}\equiv\frac{\mathrm{d}^2}{\mathrm{d} r^2}-\frac1r\frac{\mathrm{d}}{\mathrm{d} r}-k^2$, we get
\begin{equation}
    \mathsf{L}\left(\mathsf{L}+\frac{i\omega_f}{\nu}\right)\tilde\psi = 0.
    \label{eq:stream_function_final}
\end{equation}

Since both operators above commute, the general solution is the sum of the solutions of $\mathsf{L}\tilde\psi_1 = 0$ and $\left(\mathsf{L}+\frac{i\omega_f}{\nu}\right)\tilde\psi_2 = 0$, such that $\tilde\psi = \tilde\psi_1+\tilde\psi_2$. The full solution then has the form
\begin{equation}
    \tilde\psi = A\, r\,I_1(kr)+C\,r\,I_1(\kappa r),
    \label{eq:cilia_final}
\end{equation}
where $I_1$ is the modified Bessel function of the first kind and $\kappa = \sqrt{k^2-\frac{i\omega_f}{\nu}}$. The coefficients $A$ and $C$ are obtained from the slip velocity at the boundary. Note that the $K_1$ solutions are discarded to keep the flow regular at the axis. To obtain the results presented in the main text, we solve Eq.~\eqref{eq:advection_diffusion} for the odor field, using the cilia-driven velocity field generated using Eq.~\eqref{eq:cilia_final} as the advecting flow.  

\bibliography{references}

\clearpage
\widetext
\begin{center}
\textbf{\large Supporting Information: Amphibian water to land transition reveals physical limits of olfaction}
\end{center}
\setcounter{equation}{0}
\setcounter{figure}{0}
\setcounter{table}{0}
\setcounter{section}{0}
\renewcommand{\theequation}{S\arabic{equation}}
\renewcommand{\thefigure}{S\arabic{figure}}
\renewcommand{\thetable}{S\arabic{table}}
\makeatletter
\let\@hangfrom@section\saved@hangfrom@section
\let\@sectioncntformat\saved@sectioncntformat
\makeatother
\renewcommand{\thesection}{S~\Roman{section}}

\section{Asymptotics for the rate of odor capture at the walls of a cylinder or a permanent odor source at steady state}
\label{sec:si:molecules}

We again use the cylindrical approximation (Eq.~\eqref{eq:cylinder_nondim}). In the advective regime, we can neglect the axial diffusion term and the equation becomes 
\begin{equation}
    \frac{\partial \theta}{\partial T}
+
\Rey\Sc(1-Y^2)\frac{\partial \theta} {\partial Z}
=
\frac{1}{Y}\frac{\partial}{\partial Y}\left(Y\frac{\partial \theta}{\partial Y}\right).
\label{eq:si:graetz}
\end{equation}
Equation~\eqref{eq:si:graetz} is the classical Graetz model for an absorbing tube~\cite{brown1960heat,shah2014laminar}. For our estimations, we use the long and short tube limits, calculations of which are reproduced below for completeness.

The general solution to Eq.~\eqref{eq:si:graetz}  at steady state can be written in the form of the following expansion~\cite{brown1960heat,shah2014laminar},
\begin{equation}
\theta(Y,Z)=\sum_{n=1}^{\infty}A_n e^{-\lambda_n^2 \frac{Z}{\Rey\Sc}}\phi_n(Y),
\label{eq:si:graetz_series}
\end{equation}
where $\phi_n(Y)$ are radial eigenfunctions and $\lambda_n$ are the corresponding eigenvalues. In the long-tube limit, $L\gg R\Rey\Sc$, the first mode dominates and the outlet concentration decays exponentially.
The adsorption efficiency can then be approximated as
\begin{equation}
\frac{\phi_\infty
}{\phi_0}=1-0.819\exp\left(-7.3\frac{L}{R\Rey\Sc}\right).
\label{eq:si:flux_longtube}
\end{equation}
where $\phi_0=\frac{1}{2}\pi R^2Uc_0$ is the incoming odor flux at the chamber entrance. This limit is appropriate when diffusion has enough time to deplete the odorant over the chamber length, e.g.~in air. 

In the entrance-region limit, $L\ll R\Rey\Sc$, adsorption is controlled by a thin concentration boundary layer at the wall, as has been shown for the thermal boundary layer in the Graetz problem~\cite{newman1969extension}. The boundary-layer thickness, nondimensionalized by $R$, scales as
\begin{equation}
\delta(Z)=\left(\frac{9Z}{2\Rey\Sc}\right)^{1/3}.
\label{eq:si:leveque_delta}
\end{equation}
The corresponding concentration profile is
\begin{equation}
\frac{c}{c_0}=\frac{1}{\Gamma(4/3)}\int_0^{\eta}e^{-\gamma^3}\,\diff\gamma,
\qquad \eta=\frac{x}{R\delta(Z)},
\label{eq:si:leveque_profile}
\end{equation}
where $x=R-r$ denotes the distance from the wall. Integrating the wall flux gives
\begin{equation}
\phi_\infty
\approx2\pi D c_0\left(R\Rey\Sc\right)^{1/3}L^{2/3}. \nonumber
\end{equation}
Then the adsorption efficiency becomes
\begin{equation}
\frac{\phi_\infty
}{\phi_0} = 4\left[ \frac{L}{R\Rey\Sc} \right]^\frac{2}{3}.
\label{eq:si:flux_shorttube}
\end{equation}

\begin{figure}[h]
\centering
\includegraphics[width=0.6\linewidth]{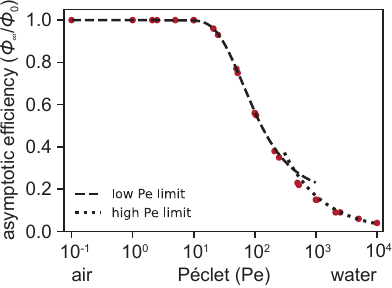}
\caption{Comparison of adsorption efficiency for the long tube (low Pe, Eq.~\eqref{eq:si:flux_longtube}) and short tube (high Pe, Eq.~\eqref{eq:si:flux_shorttube}) approximations with numerical results obtained by solving Eq.~\eqref{eq:si:graetz}, for the aquatic chamber geometry ($L/R=8.67$).}
\label{fig:si:pe_limits}
\end{figure}

A comparison of the adsorption efficiency under the long and short tube approximations with the numerical solution of Eq.~\eqref{eq:si:graetz} is shown in Fig.~\ref{fig:si:pe_limits}. 

Note that the amount of odor flowing into the chamber increases linearly with the flow rate $U$: $\phi_0 \sim c_0 \pi R^2 u$.
Because the adsorption efficiency $\phi_\infty
/\phi_0$ decreases as $U^{-2/3}$ (from
Eq.~\eqref{eq:si:flux_shorttube}), the  absolute adsorption $\phi_\infty
$ increases, albeit slowly, with flow rate (Fig.~\ref{fig:cylinder}c)
\begin{equation}
    \phi_\infty
    = 2\pi \nu c_0 \left[R\Rey\right]^\frac{1}{3}\left[\frac{L}{Sc}\right]^\frac{2}{3}.
    \label{eq:si:adsorption_limit}
\end{equation}

Table~\ref{tab:molecules} lists the odorants used to compute the saturated adsorption rates in Fig.~\ref{fig:molecules}. Figure~\ref{fig:molecules} features the hydrodynamic adsorption efficiency, $\phi_\infty
$, assuming $c_0$ is the saturated water and air concentrations. 
Saturated water concentrations are taken from reported water solubilities and saturated air concentrations are computed from reported vapor pressures using the ideal gas law. Values are obtained from PubChem~\cite{kim2025pubchem} or The Good Scents Company~\cite{goodscents}. To evaluate the adsorption efficiencies $\phi_\infty$ from a constant source (Fig.~\ref{fig:molecules}a), we make use of the approximations given by Eqs.~\eqref{eq:si:flux_longtube} and \eqref{eq:si:flux_shorttube} in air and water, respectively, given their excellent agreement with the numerical results (Fig.~\ref{fig:si:pe_limits}). The approximations require $L$, Re and Sc for each molecule. Consistent with our experimental results, we choose $L = 4.6$\,mm. Reynolds number $\Rey = \frac{UR}{\nu}$, where $U$ is the maximum fluid velocity inside the olfactory chamber. We choose Re to be $1$, with the error bars representing the range $0.1<$ Re $<10$. We deliberately choose a non zero value for aquatic inhalation although the measured value is negligible (Fig.~\ref{fig:pulse}e). This helps us estimate whether, even with inhalation, aquatic olfaction delivers sufficient odorants to the epithelium. Informed by the experimental results, we choose $R = 0.53$\,mm. We choose Sc to be 1 in air and 1000 in water. To measure $\phi_\infty
$ from an ephemeral odor source, we consider a saturated odor pulse of radius $R/2$. We then evaluate the adsorption rate from the odor flux towards the epithelium (Fig.~\ref{fig:si:pulse_flux}). The reported values are obtained from the average of the odor flux at Re$=0$ and the maximum odor flux.

\section{Short time limit of adsorption inside a cylinder for an ephemeral odor source}
\label{sec:si:short_time_limit}

Consider an instantaneous whiff of odor placed at the center of a long cylinder with radius $R$, represented by a Dirac delta function. The source contains $N_0$ moles of odor that are released at time $t=0$. Since we are interested in the diffusive transfer towards the wall, we consider no flow inside the cylinder. The diffusivity is $D$ and the walls of the cylinder are perfectly adsorbing. We are interested in the short-time limit of the cumulative 
adsorbed odor $N(t)$ at time $t\ll \frac{R^2}{D}$. 

Since axial and radial diffusion are independent and axial diffusion does not affect the probability of reaching the cylindrical wall, this problem simplifies to adsorption inside a 2D disk. In the absence of the wall, the diffusion equation has a solution for the two-dimensional concentration of odor, $c_{\mathrm{free}}$,  of the form:
\begin{equation}
    c_{\mathrm{free}} = \frac{N_0}{4\pi Dt}e^{\left( -\frac{r^2}{4Dt} \right)}
\end{equation}
inside the disk, where $r$ is the radial coordinate. 

\begin{figure}[h]
\centering
\includegraphics[width=0.5\linewidth]
{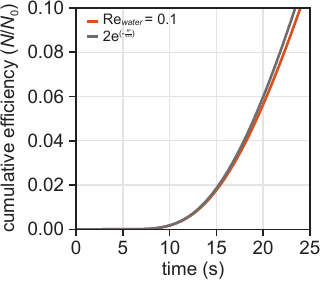}
\caption{Comparison of the short-time limit of adsorption inside a cylinder with the numerical results for adsorption of a pulse source in water with $\Rey = 0.1$.}
\label{fig:si:adsorption_short_time}
\end{figure}

The fraction of odor lying outside $R$ is then given by
\begin{equation}
    \frac{1}{N_0}\int_R^\infty 2\pi r c_{\mathrm{free}} \diff r = e^{\left( -\frac{R^2}{4Dt} \right)}.
    \label{eq:si:prob_free}
\end{equation}
Equation~\eqref{eq:si:prob_free} is also the probability $P_\mathrm{free}$ that a freely diffusing particle lies outside the disk at time $t$. However, since we have a perfectly adsorbing wall, we are interested in the cumulative probability of first hitting the wall by time $t$, $P_\mathrm{hit}$.  
For a planar wall, an odor particle that has first reached the wall before time $t$ has an equal probability of wandering on either side of the wall, and being found 
inside or outside the wall at time $t$, and thus $P_\mathrm{hit} = 2P_\mathrm{free}$ 
Although the wall is curved, at short times diffusion probes only a thin layer near the wall, over which its curvature is negligible and the wall can be treated as locally planar. Thus we assume this relation to hold.
The short time, cumulative adsorbed odor $N$ is then given by
\begin{equation}
    N (t) \approx 2N_0 e^{\left( -\frac{R^2}{4Dt} \right)}.
    \label{eq:si:adsorption_short_time}
\end{equation}
as presented in the main text (Eq.~\eqref{eq:adsorption_short_time}). Fig.~\ref{fig:si:adsorption_short_time} compares this limit with the data presented in Fig.~\ref{fig:pulse}b, showing that the small flows observed in aquatic axolotls do not affect odor capture sensibly.

\section{Sensing timescale for ephemeral odor whiffs}
\label{sec:si:timescale_adsorption_limits}
\begin{figure}[h]
\centering
\includegraphics[width=0.9\linewidth]{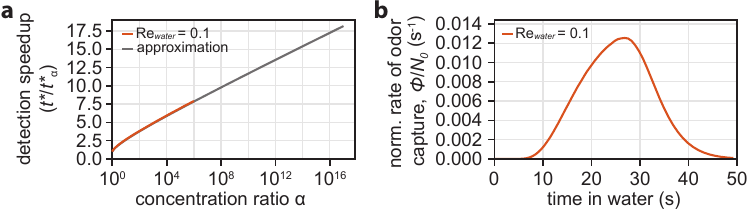}
\caption{{\bf a}, Speedup in odor detection from a pulse source, $t^*/t_\alpha^*$, when the inlet concentration is increased by a factor $\alpha$, with detection defined by a threshold on the instantaneous adsorbed flux. The orange curve is the numerical solution with Re = 0.1. The threshold when $\alpha = 1$ is the maximum flux in the numerical solution. 
{\bf b}, Total odor flux into the wall of a cylinder in water for an instantaneous odor pulse located at the inlet.}
\label{fig:si:flux_limit}
\end{figure}

Here we evaluate how the timescale of detection of an ephemeral odor whiff 
changes as a function of the concentration of odor in the whiff. In contrast to Sec.~\ref{sec:si:short_time_limit} and the corresponding discussion in the main text, where detection is set by the cumulative amount of odor adsorbed, we here adopt a threshold on the instantaneous adsorbed flux. Which of the two is the relevant criterion depends on how olfactory receptors integrate their input: a flux threshold is arguably the more natural choice if detection is triggered by the instantaneous rate at which odorant reaches the epithelium rather than by an accumulated dose. We will use our measurements in axolotls to estimate the threshold flux needed to trigger a response.  
Starting from Eq.~\eqref{eq:si:adsorption_short_time}, the short-time limit for the adsorbed odor flux for an ephemeral whiff of odor, $\phi(t)=dN/dt$, is
\begin{equation}
    \phi(t)\approx \frac{N_0t_D}{2t^2} e^{\left( -\frac{t_D}{4t} \right)} , \qquad t_D = \frac{R^2}{D}.
    \label{eq:si:adsorption_flux_short_time}
\end{equation}
The flux is not monotonic in time. We therefore define the detection time $t^*$ as the first time at which $\phi$ reaches a threshold $\phi_t$, that is, the crossing on the rising branch. Equation~\eqref{eq:si:adsorption_flux_short_time} cannot be inverted in elementary form, so we obtain $t^*$ numerically. Figure~\ref{fig:si:flux_limit} shows that as the concentration within the whiff, and thus $N_0$, increases by a factor $\alpha$, the time to reach a fixed threshold in the flux $\phi$ can be greatly reduced. However, the speed up is approximately logarithmic, hence it requires considerably more concentrated sources.

\section{Triggering flux from in vitro dose response curves}
\label{sec:J50}
In order to estimate if odor may trigger a response in vivo, i.e.~observed in behavior, 
we first use our dose response curves \emph{in vitro} to estimate how many molecules of odor must hit the epithelium to trigger a response. \\
In our experiments, mammalian cells expressing axolotl receptors were activated in vitro when they were flushed with odor at the half maximal effective concentrations (EC$_{50}$) that ranged from $c_{50}=80\,n$M to 800$\,\mu$M, depending on receptor and odorant. To estimate the flux corresponding to these concentrations, we model the process that leads odor molecules to diffuse to the receptors in these experiments.
The experiments are conducted in 96 well plates; cells cover at least 95\% of the bottom surface. Modeling the cells as a perfectly absorbing surface located at $z=0$, we estimate the flux of molecules hitting the receptors by solving the diffusion equation
$$c_t(\mathbf{r},t)=Dc_{zz}(\mathbf{r},t)$$ 
with initial condition $c(\mathbf{r},t=0)=c_{50}\, \forall \, \mathbf{r}$ and absorbing boundary condition at $z=0$ i.e.~$c(\mathbf{r},t)_{z=0}=0\,\forall x,y,t$. The solution is 
$$c(\mathbf{r},t)=\frac{c_{50}}{2}\left(\text{erfc}\left(\frac{z}{2\sqrt{Dt}}\right)- \text{erfc}\left(\frac{-z}{2\sqrt{Dt}}\right)\right).$$ 
The local flux toward the absorbing boundary per unit time and surface is therefore 
$J=-D\partial_zc|_{z=0} = c_{50}\sqrt{D/(\pi t)}$, which depends on time and averages to 
\begin{equation}
\label{eq:flux_trigger}
J_{50}=\frac{1}{T}\int_0^T J(t)dt = 2c_{50}\sqrt{\frac{D}{\pi T}}
\end{equation}
after a time $T$. 
Equation~\eqref{eq:flux_trigger} computed at the onset of the response,  $T$
and at the threshold EC$_{50}$ concentration, $c_{50}$, represents the triggering flux, i.e.~the minimum number of molecules of odor that must reach the epithelium per unit time and surface in order to trigger a response. 
In reality, odor is not delivered instantaneously, but it is delivered to the 96 well plates through a pipette, where we assume it mixes homogeneously. Activation of the receptors is monitored by a slow build up in luminescence, over a timescale of several seconds. Hence we use equation~\eqref{eq:flux_trigger} at $T=T_{\text{mix}}=1$~s to estimate the triggering flux. 
Using $D\approx 10^{-5}$cm$^2$/s and EC$_{50}$ concentrations $c_{50}=10^{-7}$ to $10^{-3}$ M we obtain an estimate of the triggering flux ranging from $J_{50}=4 \times 10^{-4} \text{ to 4 nmol/cm}^2 s$.

The cumulative flux of molecules reaching the unit surface in time $T$ is 
\begin{equation}
\label{eq:cumulative_local_triggering_flux}
\sigma_{50}(T)=\int_0^T J(t)dt = 2c_{50}\sqrt{\frac{DT}{\pi}}.
\end{equation}
Using the same parameters as above, the triggering cumulative flux  ranges from $\sigma_{50}\approx 3\times 10^{-13}$ to $3\times 10^{-9}$~mol/cm$^2$.\\ 

\section{Can an ephemeral whiff trigger a behavioral response?}
\begin{figure}
\includegraphics[width=\linewidth]{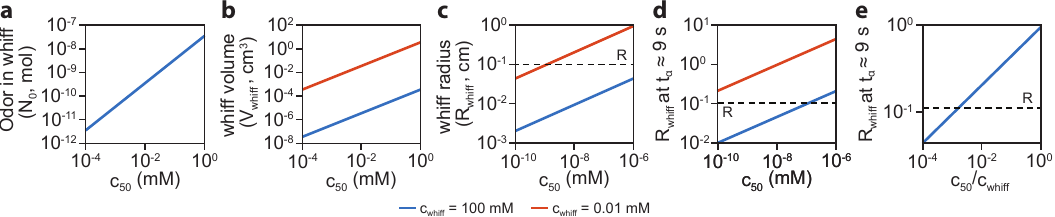}
\caption{
{\bf a}, Minimum amount of odor molecules $N_0$ in an ephemeral whiff that triggers a response, when the EC$_{50}$ threshold for response $c_{50}$ ranges from $10^{-4}$~mM to $1$~mM. 
{\bf b}, Volume and {\bf c}, equivalent radius of such whiff when it contains saturated highly soluble odor ($c_{\text{whiff}}=100$~mM, solid line)
or weakly soluble odor ($c_{\text{whiff}}=0.01$~mM, dashed line). 
{\bf d}, Same as {\bf c}, but ensuring that the response is triggered in about 10~s, corresponding to a threefold speed up obtained at $\alpha=100$. 
{\bf e}, Same as {\bf d}, but plotted as a function of the ratio between in vitro $c_{50}$ and concentration in the whiff, $c_{\text{whiff}}$. The dashed line in {\bf c}-{\bf e} marks a reference size of 0.1~cm matching the aperture of the nasal cavity.}
\label{fig:minimumN}
\end{figure}

Can an ephemeral whiff of odor provide the triggering flux of odor to activate a response? And how long does it take to build up the triggering flux? From Eq.~\eqref{eq:si:adsorption_flux_short_time}, and assuming odor is captured homogeneously along the whole chamber, a whiff of odor generates an average flux of molecules per unit surface and unit time:
$$J(t)=\frac{\phi(t)}{2\pi RL}\approx \frac{N_0R}{4\pi DLt^2}e^{-t_D/4t}.$$
Its maximum value is achieved at $t=t_D/8$ and is \begin{equation}
\label{eq:max_flux}
J_{\text{max}}=16 N_0D/(\pi LR^3) e^{-2}.
\end{equation}
The maximum flux of molecules hitting the epithelium within the axolotl nasal cavity (Eq.~\eqref{eq:max_flux}) reaches the triggering flux for activation (Eq.~\eqref{eq:flux_trigger}) if 
\begin{equation}
\label{eq:condition}
N_0>\sqrt{\frac{\pi}{DT}}\frac{LR^3}{8e^{-2}}c_{50}.
\end{equation}
The amount of odor required to trigger a response from Eq.~\eqref{eq:condition} is plotted in Fig.~\ref{fig:minimumN}(a) using $R = 0.53$\,mm, $L/R=8.67$, $D=10^{-5} cm^2/s$, $T=1s$. 
This condition can be achieved at high concentrations, which may be relevant if the animal is close to the target. Indeed, to contain the required $N_0$ moles of odor to trigger a response, a whiff would need to have volume $V_{\text{whiff}}=N_0/c_{\text{whiff}}$ (Fig.~\ref{fig:minimumN}(b)) and equivalent radius $R_{\text{whiff}}=(V_{\text{whiff}}/(4/3\pi))^{1/3}$ (Fig.~\ref{fig:minimumN}(c)). For  $c_{\text{whiff}}$ we used the minimum (0.01 mM, red dashed line)  and maximum (100 mM, blue solid line) saturated concentrations estimated over a panel of odorants (Table~\ref{tab:molecules}). In most conditions, the whiff is much smaller than the nose itself (see~Fig.~\ref{fig:minimumN}b-c). Hence, in conclusion, individual whiffs at saturated concentration may trigger a response. \\

Next we ask how long it takes to trigger the response. The timescale to reach peak adsorption is $t^*=t_D/8 =  R^2/(8D)\approx 35 s$; the response is triggered at a shorter timescale $t^*_\alpha$ if the peak flux is  larger than the flux needed to trigger the response:
\begin{equation}
\label{eq:tstar_alpha}
t^*_\alpha = \frac{R^2}{8D}\frac{t^*_\alpha}{t^*} \quad \text{ where } \quad\alpha=\frac{J_{\text{max}}}{J_{50}}\gg 1.
\end{equation}
The speedup $\frac{t^*_\alpha}{t^*}$ can be read directly from Fig.~\ref{fig:si:flux_limit}. For example, in order to trigger a response in about 10~s, we need a threefold speedup. From Fig.~\ref{fig:si:flux_limit} we find that at $\alpha \approx 100$ the speedup is about 3.75, leading to a response timescale $t^*_\alpha\approx 9$~s. The condition $\alpha=100$ is easily achieved at saturated concentrations for our most soluble odor, but not for our least soluble odor, which requires extremely large whiffs even larger than the aperture of the nasal cavity (see~Fig.~\ref{fig:minimumN}d). In practice, the odor must be at least 1000 times more concentrated than the $c_{50}$ concentration that triggers a response \emph{in vitro} (see~Fig.~\ref{fig:minimumN}e). Odor whiffs that are sensed well (low $c_{50}\approx 90$~nM) trigger a response at $\approx 90\,\mu$M; but odors sensed poorly (high $c_{50}\approx 1$~mM) may never trigger a response.  

Cilia may provide further speed up: for example, metachronal pumping contributes a further $\sim 3$ fold speedup at $\lambda/R = 2$. 
In the absolute best scenario, the largest possible whiff (equivalent radius $\sim 2R \approx  0.1$~cm), of the most soluble odor at saturation, $c_{\text{whiff}}=100$~mM, would contain $N_0=4\times 10^{-7}$ moles of odor; our best sensed odors had \emph{in vitro} half maximal activation threshold $c_{50}=10^{-7}$\,M the number of moles required for activation is 10$^{-12}$, thus $\alpha=10^5$, providing a speedup of a factor $\sim$~7. With an additional three-fold speedup from metachronal waves, the response could start as soon as 1.7 seconds. Thus, a whiff of concentrated odor may trigger a response. However, even in the most favorable conditions with an odor near saturation, sensed well by the animal in the presence of ciliary pumping, the response may be triggered in a timescale of seconds: still thousands of times slower than in air.

\section{Can a large persistent odor patch trigger a behavioral response?}
\begin{figure}[h]
\centering
\includegraphics[width=0.5\linewidth]{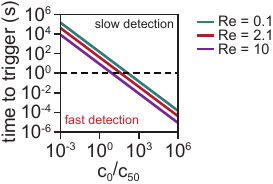}
\caption{Extended odor patches near saturation may trigger a behavioral response in water. The figure shows the timescale to trigger a response as a function of relative concentration $c_0/c_{50}$ from Eq.~\eqref{eq:timescale_constant}. Colors indicate different Reynolds numbers (Re=0.1, 2.1 and 10, corresponding to speeds of 0.2 mm/s, 4 mm/s and 19 mm/s respectively).
}
\label{fig:trigger_constant_source}
\end{figure}

In our laboratory conditions, a large persistent odor patch is directly pipetted at the animals' nose. 
We now show that a persistent odor patch may trigger a much faster response if inhaled at a finite speed. To model this case, we consider a permanent, extended odor source at concentration $c_0$, and an inflow with constant speed $U$ bringing odor into the nasal cavity. 
The time dependent problem with a fixed concentration and speed at the inlet was studied numerically and asymptotically in the main text and in section~\ref{sec:si:molecules}. In a nutshell, because of the inefficiencies of diffusion in water, 
the rate of molecules reaching the entire epithelium per unit time ($\phi$) is much lower than the rate of odor flowing into the cavity ($\phi_0$). The efficiency of odor adsorption, $\epsilon=\phi/\phi_0$ varies with the Peclet number and with time
(Fig.~\ref{fig:cylinder}(d)) and it reaches a maximum value that can be approximated in water by Eq.~\eqref{eq:si:flux_shorttube}. Hence
$$\phi(t)=\epsilon(t)\phi_0<\epsilon_\infty \phi_0$$
\noindent where the incoming odor flux $\phi_0$ is constant, as long as the source is persistent over the timescale of interest, $\phi_0 = c_0 \pi R^2 U$. 
If we assume that odor flows to the entire epithelium homogeneously (a reasonable assumption in water, as seen in main Fig.~\ref{fig:cfd}), the amount of odor adsorbed per unit time and surface is:
\begin{equation}
\label{eq:psi_constant_t}
J(t) \approx  \frac{1}{2\pi RL}\phi(t)
=\frac{c_0RU}{2 L}\epsilon
\end{equation}
\noindent 
and asymptotically, the flux  saturates to:
\begin{equation}
\label{eq:psi_constant}
J_\infty = \frac{c_0RU}{2 L}\epsilon_\infty
=2 c_0\left(\frac{UD^2}{RL}\right)^{1/3}
\end{equation}
\noindent 
where we used $\epsilon_\infty$ from Eq.~\eqref{eq:si:flux_shorttube}.
To trigger a response, $J_\infty$  (Eq.~\ref{eq:psi_constant}) must exceed the triggering flux $J_{50}$ (Eq.~\eqref{eq:flux_trigger}), which imposes that the concentration should satisfy:
\begin{equation}
\label{eq:psimax_const}
\left(\frac{c_{0}}{c_{50}}\right)^2>\frac{t_U^{2/3}t_D^{1/3}}{\pi T}
\end{equation}
where $t_U=L/U$ is the advective timescale and, as above, $t_D=R^2/D$ is the diffusive timescale. We use values from our experiments in axolotls : $R = 0.53$\,mm, $L/R=8.67$, $D=10^{-5} cm^2/s$,  triggering EC50 concentrations ranging from $c_{50}=10^{-4}$~mM to 1~mM and the mixing timescale $T\approx 1$~s. The result is that asymptotically (at long times), odor capture from extended permanent sources are sufficient to trigger a response if they are about 4 times more concentrated than their \emph{in vitro} EC$_{50}$ (at Re=0.1) or about as concentrated as their \emph{in vitro} EC$_{50}$. 
\\ 
But this flux of odor is only reached asymptotically i.e.~after long enough. Next, we show that the response may kick in on timescales that range widely from ms to minutes. To quantify the response timescale, we approximate $\epsilon$ linearly, $\epsilon\approx \dot\epsilon_{Re} t$ and we match the actual flux $J(t)$ (Eq.~\eqref{eq:psi_constant_t}) to the required flux to trigger a response $J_{50}$ (Eq.~\eqref{eq:flux_trigger}), which yields:
\begin{equation}
\label{eq:timescale_constant}
t>\frac{4}{\sqrt{\pi}}\frac{c_{50}}{c_0} \frac{t_U}{\dot\epsilon}(t_{mix}t_D)^{-1/2}
\end{equation}
\noindent where as above $t_D=R^2/D$; $t_U =L/U$; $t_{mix}\approx 1 \, s$ and we use $\dot\epsilon(Re>2.1)\approx 0.0038$~s$^{-1}$ and $\dot\epsilon(Re=0.1)\approx 0.0223 $~s$^{-1}$, with the linear approximation being valid at times shorter than 2 s. From Eq.~\eqref{eq:timescale_constant} we obtain timescales ranging widely depending on water flow, dictating the Reynolds number; on how concentrated the odor source is ($c_0$) and how well the odor is sensed ($c_{50}$), 
see Fig.~\ref{fig:trigger_constant_source}, right. To ensure a response within about 1~s, matching our experimental observations, we find that odor concentration must be much larger than the concentration that triggers a response \emph{in vivo}, i.e.~$c_{0}\gtrsim 100 c_{50}$. 
Our \emph{in vitro} $c_{50}$ ranged from 0.1~$\mu$M to 1~mM; 
for these odors a behavioral response under 1~s would be possible for patches ranging from 10$\mu$M to 100~mM, which is a relatively high concentration considering that saturated concentrations for our panels of odors ranged from 0.01~mM to 100 mM. 
Results are shown for different Reynolds numbers (Red: Re = 0.1; blue Re = 2.1; green: Re = 10). The highest Re = 10, corresponding to incoming flows of almost 2~cm/s is unrealistic for natural breathing by aquatic axolotls. Indeed, we could not make any reliable measurements; hence the flow could only be below the sensitivity of our probe. It is however still relevant for experiments, where odor is introduced with a pipette which creates flow exogenously. 

\section{Behavioral responses to odor introduced in the water tank}
\label{sec:model_behavior}
Our behavioral experiments (Fig.~6~main text) show that when odor is introduced into the tank, it elicits behavioral responses over timescales that can be as fast as 0.4 seconds, but only at relatively high concentrations. 

\paragraph{Odor delivery, $t\lesssim 1s$.} While odor is delivered, the flux of odor molecules captured at the olfactory epithelium per unit time and surface is expected to increase linearly with time. To estimate how odor capture increases with time, we proceed to quantify how odor is delivered. \\
1~mL of odor is pipetted into the water tank at a distance $x= 0.8\pm0.5$~cm from the nose, at an angle close to normal to the nose. Odor delivery takes a timescale of $1.1\pm0.3$~s. Visually, it appears that for the duration of odor delivery, the odor in the jet is not diluted. 
The diameter of the pipette at the tip is 0.8~mm. While the odor is pipetted towards the animal, it creates a turbulent jet (speed at the tip $U\sim V_{tip}/\pi r_{tip}^2\approx 1.7$~m/s Reynolds number $Re\approx 8000$). Once the jet reaches the animal, the centerline speed reduces to $U_{animal}\sim U_{tip}Br_{tip}/x \sim 80 cm/s$. 
At distance $x\approx 0.8$~cm, the jet has not spread much beyond $r_{tip}$ (half width $r_{1/2}\approx S x$, with $S\approx 0.1$). 
The jet centerline hits the animal at a distance $y=2.5\pm2.9$~mm from the nostril. Because the jet remains about 0.4~mm wide, it does not engulf the nostril. As the jet hits the animal and spreads along the snout, it enters the nostril and the olfactory chamber by residual velocity fluctuations. In circular jets, lateral fluctuations are $\sim2\%$ of the centerline velocity~\cite{pope2000turbulent}, i.e.~$U_{nose}\approx 1.6$~cm/s. The flux of odor molecules during odor delivery in our behavioral experiments can be obtained from our theory for extended source at finite speed (Eq.~\eqref{eq:psi_constant_t}), where $U=U_{nose}$, $\epsilon\approx \dot\epsilon t $, $\dot \epsilon\sim 0.0038 s^{-1}$, the radius and length of the axolotls and undiluted concentrations from $c_0=10$~nM to $c_0=1$~mM, used in these experiments (Fig.~\ref{fig:experimental}c).\\

\paragraph{After odor delivery, $t\gg1 s$.}
Once the odor is delivered, the pipette is extracted from the tank, the water slows to rest and odor is diluted 1000 times (1mL odor in 1L water in tank). Visually, this appears to happen across several seconds. The water speed then drops to zero and the concentration is 1000 times diluted. 
Regardless of the details of this regime, the amount of odor molecules captured by the epithelium per unit time and surface cannot increase. Hence, if the odor was not sensed while it was delivered, it cannot be sensed afterwards. 

Figure~\ref{fig:experimental}c shows that under this protocol for olfactory stimulation, only the three highest concentrations reach the triggering flux, matching our experimental observations. 

\section{Effect of boundary conditions}
\label{sec:si:boundary_condition_effect}
A refined argument would be needed to account for imperfect adsorption both in the \emph{in vitro} heterologous system as well as the predictions for flux toward the epithelium within the nasal cavity. Indeed, molecules reflected by the surface would greatly increase the number of collisions with the surface, as it is well known that diffusion leads molecules to probe surfaces many times. 
A more detailed treatment of this problem requires further modeling of unknown quantities like the binding efficiency and the spatial distribution of receptors, hence are not accessible at this time. 
We expect perfect absorption to be a reasonable approximation for the \emph{in vitro} experiments, because cells are extremely crowded and overexpress the receptor.
\emph{In vivo}, identifying the correct boundary condition requires a careful modeling of the process of odor binding, including the efficiency of the odor transport proteins in the mucus layer that covers the epithelium; the coverage of different receptors across the epithelium and the complex landscape of binding efficiencies of receptor-ligand pairs. Note that because random walkers dwell near surfaces before escaping away, even if a very small fraction of the epithelium is covered by the receptors, the surface remains nearly perfectly absorbing (for a spherical adsorber a 10$^{-4}$ surface coverage affords nearly 50\% adsorption~\cite{berg2025random}).
\\

\begin{figure}[h]
\centering
\includegraphics[width=0.8\linewidth]{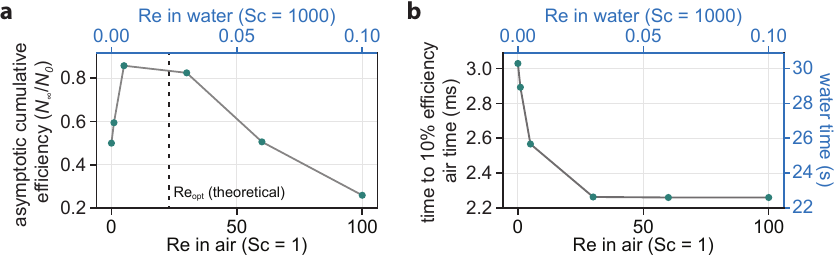}
\caption{Adsorption of an odor pulse located at the inlet of a cylinder (same as Fig.~2c,d, as a function of $\Rey$). {\bf a}, Asymptotic adsorption efficiency $N_\infty/N_0$ as a function of Re with the lower axis giving Re in air ($\Sc=1$) and the upper axis Re in water ($\Sc=1000$). In water, $N_\infty/N_0$ achieves a maximum at very low Re. {\bf b}, Time to reach $N/N_0 = 0.1$ of the total odor as a function of Re. Even for the optimal Re, the time to reach $N/N_0 = 0.1$ takes over 20\,s in water.}
\label{fig:si:pulse}
\end{figure}

\begin{figure}[h]
\centering
\includegraphics[width=0.6\linewidth]{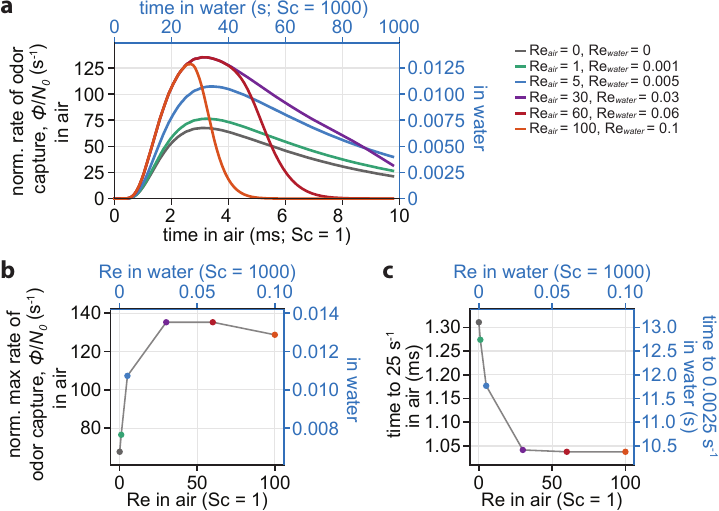}
\caption{Total odor flux into the walls of a cylinder for an instantaneous odor pulse, as a function of $\Rey$. {\bf a}, Odor flux as a function of time for different values of $\Rey$. {\bf b}, Maximum odor flux as a function of $\Rey$. The maximum increases with $\Rey$, eventually reaching a peak at an intermediate $\Rey$. {\bf c}, Time to reach a threshold flux as a function of $\Rey$.}
\label{fig:si:pulse_flux}
\end{figure}

\begin{table*}[t]
\caption{Panel of odor molecules used to estimate saturated water and air adsorption rates.}
\label{tab:molecules}
\begin{ruledtabular}
\scriptsize
\begin{tabular}{l c c c}
Name & Saturated aqueous concentration (mol/L) & Saturated gas-phase concentration (mol/L) & Air or water sensed \\
\hline
Ethyl tiglate        & $4.71\times 10^{-2}$  & $2.30\times 10^{-4}$ & Air   \\
Allyl tiglate        & $4.09\times 10^{-3}$  & $6.83\times 10^{-5}$ & Air   \\
Hexyl tiglate        & $8.25\times 10^{-5}$  & $8.07\times 10^{-6}$ & Air   \\
Methyl tiglate       & $3.34\times 10^{-2}$  & $7.20\times 10^{-4}$ & Air   \\
Isopropyl tiglate    & $3.50\times 10^{-3}$  & $1.01\times 10^{-4}$ & Air   \\
Citronellyl tiglate  & $1.62\times 10^{-6}$  & $1.97\times 10^{-7}$ & Air   \\
Benzyl tiglate       & $3.25\times 10^{-4}$  & $4.30\times 10^{-7}$ & Air   \\
Phenylethyl tiglate  & $3.23\times 10^{-4}$  & $4.19\times 10^{-8}$ & Air   \\
2-Ethylhexanal       & $3.12\times 10^{-3}$  & $9.68\times 10^{-5}$ & Air \\
Benzyl acetate       & $2.06\times 10^{-2}$  & $9.14\times 10^{-6}$ & Air   \\
Salicylic acid       & $1.63\times 10^{-2}$  & $4.41\times 10^{-9}$ & Air   \\
Phenylacetic acid    & $1.27\times 10^{-1}$  & $2.04\times 10^{-7}$ & Air \\
4-Allyl anisole      & $1.20\times 10^{-3}$  & $2.69\times 10^{-6}$ & Air \\
Ethyl valerate       & $1.70\times 10^{-2}$  & $2.55\times 10^{-4}$ & Air   \\
Citronellal          & $2.52\times 10^{-4}$  & $1.51\times 10^{-5}$ & Air   \\
(+)-Carvone          & $8.65\times 10^{-3}$  & $8.60\times 10^{-6}$ & Air   \\
($-$)-Carvone        & $8.65\times 10^{-3}$  & $8.60\times 10^{-6}$ & Air   \\
2-Methoxypyrazine    & $3.38\times 10^{-2}$  & $2.28\times 10^{-4}$ & Air   \\
Isoeugenol           & $4.93\times 10^{-3}$  & $1.08\times 10^{-6}$ & Air \\
Methyl valerate      & $4.36\times 10^{-2}$  & $9.84\times 10^{-4}$ & Air   \\
Acetophenone         & $5.10\times 10^{-2}$  & $2.10\times 10^{-5}$ & Air   \\
Phenyl acetate       & $3.41\times 10^{-2}$  & $2.14\times 10^{-5}$ & Air   \\
Methyl benzene       & $5.71\times 10^{-3}$  & $1.53\times 10^{-3}$ & Air   \\
Methyl salicylate    & $4.60\times 10^{-3}$  & $1.61\times 10^{-6}$ & Air   \\
Nonyl acetate        & $5.83\times 10^{-5}$  & $1.06\times 10^{-5}$ & Water \\
Octanoic acid        & $6.91\times 10^{-3}$  & $2.00\times 10^{-7}$ & Water \\
Pentanol             & $2.50\times 10^{-1}$  & $1.18\times 10^{-4}$ & Water   \\
Heptyl acetate       & $6.44\times 10^{-4}$  & $2.69\times 10^{-5}$ & Water \\
Hexyl acetate        & $4.47\times 10^{-3}$  & $7.10\times 10^{-5}$ & Water \\
Undecanoic acid      & $2.80\times 10^{-4}$ & $8\times 10^{-9}$  & Water \\
Nonanol              & $1.23\times 10^{-3}$  & $1.08\times 10^{-6}$ & Water   \\
Hexanol              & $5.77\times 10^{-2}$  & $4.95\times 10^{-5}$ & Water \\
Decanoic acid        & $3.59\times 10^{-4}$  & $1.97\times 10^{-8}$ & Water \\
Decanol              & $2.34\times 10^{-4}$  & $4.58\times 10^{-7}$ & Water   \\
Octyl acetate        & $1.94\times 10^{-4}$  & $2.15\times 10^{-5}$ & Water \\
Heptanoic acid       & $1.86\times 10^{-2}$  & $5.38\times 10^{-7}$ & Water \\
Decyl acetate        & $1.76\times 10^{-5}$  & $1.87\times 10^{-7}$ & Water \\
Heptanol             & $1.44\times 10^{-2}$  & $1.13\times 10^{-5}$ & Water \\
Butanol              & $8.53\times 10^{-1}$  & $3.76\times 10^{-4}$ & Water \\
Nonanoic acid        & $1.79\times 10^{-3}$  & $8.87\times 10^{-8}$ & Water \\
Hexanoic acid        & $8.87\times 10^{-2}$  & $2.15\times 10^{-6}$ & Water \\
Ethyl propionate     & $1.88\times 10^{-1}$  & $1.93\times 10^{-3}$ & Water \\
Propyl acetate       & $1.85\times 10^{-1}$  & $1.93\times 10^{-3}$ & Water \\
Isobutyl propionate  & $8.22\times 10^{-3}$  & $3.92\times 10^{-4}$ & Water \\
Allyl butyrate       & $9.62\times 10^{-3}$  & $2.39\times 10^{-4}$ & Water \\
Methyl propionate    & $7.08\times 10^{-1}$  & $4.52\times 10^{-3}$ & Water \\
Pentyl acetate       & $1.33\times 10^{-2}$  & $1.88\times 10^{-4}$ & Water \\
Valeric acid         & $2.35\times 10^{-1}$  & $1.02\times 10^{-5}$ & Water \\
Octanal              & $4.37\times 10^{-3}$  & $6.35\times 10^{-5}$ & Water   \\
2-Hexanone           & $1.72\times 10^{-1}$  & $6.24\times 10^{-4}$ & Water \\
Methyl butyrate      & $1.47\times 10^{-1}$  & $1.74\times 10^{-3}$ & Water \\
2-Heptanone          & $3.75\times 10^{-2}$  & $2.07\times 10^{-4}$ & Water \\
Butyl acetate        & $7.17\times 10^{-2}$  & $6.18\times 10^{-4}$ & Water \\
Valeraldehyde        & $1.36\times 10^{-1}$  & $1.40\times 10^{-3}$ & Water \\
Isoamyl acetate      & $1.54\times 10^{-2}$  & $2.70\times 10^{-4}$ & Water \\
\end{tabular}
\end{ruledtabular}
\end{table*}

\end{document}